\documentclass[10pt, conference, letterpaper]{ IEEEtran}
\IEEEoverridecommandlockouts
\usepackage{cite}
\usepackage{amsmath,amssymb,amsfonts}
\usepackage{algorithmic}
\usepackage{graphicx}
\usepackage{textcomp}
\usepackage{xcolor}
\usepackage{threeparttable}

\usepackage{amsmath}
\usepackage{amsthm}
\usepackage{multirow}
\usepackage{subfigure}
\usepackage{enumitem}
\usepackage[ruled,vlined]{algorithm2e}  
\usepackage{ragged2e} 
\usepackage{booktabs,makecell, multirow, tabularx}

\def\BibTeX{{\rm B\kern-.05em{\sc i\kern-.025em b}\kern-.08em
    T\kern-.1667em\lower.7ex\hbox{E}\kern-.125emX}}
\begin{document}

\title{Seer: Proactive  Revenue-Aware Scheduling \\
for Live Streaming Services in\\
 Crowdsourced Cloud-Edge Platforms}
% {\footnotesize \textsuperscript{*}Note: Sub-titles are not captured in Xplore and
% should not be used}
% \thanks{Identify applicable funding agency here. If none, delete this.}

\author{\IEEEauthorblockN{Shaoyuan Huang\IEEEauthorrefmark{2}, Zheng Wang\IEEEauthorrefmark{2}, Zhongtian Zhang\IEEEauthorrefmark{2}, Heng Zhang\IEEEauthorrefmark{2}, Xiaofei Wang\IEEEauthorrefmark{2}\IEEEauthorrefmark{1}, Wenyu Wang\IEEEauthorrefmark{4}}
\IEEEauthorblockA{\IEEEauthorrefmark{2}College of Intelligence and Computing, Tianjin University, Tianjin, China}
\IEEEauthorblockA{\IEEEauthorrefmark{4}PPIO Cloud Computing (Shanghai) Co., Ltd., Shanghai, China}
\IEEEauthorblockA{\{hsy\_23, wz\_424, 3021210045, hengzhang, xiaofeiwang\}@tju.edu.cn, wayne@pplabs.org}
\thanks{\IEEEauthorrefmark{1}Xiaofei Wang is the corresponding author.}
}

% \author{\IEEEauthorblockN{1\textsuperscript{st} Given Name Surname}
% \IEEEauthorblockA{\textit{dept. name of organization (of Aff.)} \\
% \textit{name of organization (of Aff.)}\\
% City, Country \\
% email address or ORCID}
% \and
% \IEEEauthorblockN{2\textsuperscript{nd} Given Name Surname}
% \IEEEauthorblockA{\textit{dept. name of organization (of Aff.)} \\
% \textit{name of organization (of Aff.)}\\
% City, Country \\
% email address or ORCID}
% \and
% \IEEEauthorblockN{3\textsuperscript{rd} Given Name Surname}
% \IEEEauthorblockA{\textit{dept. name of organization (of Aff.)} \\
% \textit{name of organization (of Aff.)}\\
% City, Country \\
% email address or ORCID}
% \and
% \IEEEauthorblockN{4\textsuperscript{th} Given Name Surname}
% \IEEEauthorblockA{\textit{dept. name of organization (of Aff.)} \\
% \textit{name of organization (of Aff.)}\\
% City, Country \\
% email address or ORCID}
% \and
% \IEEEauthorblockN{5\textsuperscript{th} Given Name Surname}
% \IEEEauthorblockA{\textit{dept. name of organization (of Aff.)} \\
% \textit{name of organization (of Aff.)}\\
% City, Country \\
% email address or ORCID}
% \and
% \IEEEauthorblockN{6\textsuperscript{th} Given Name Surname}
% \IEEEauthorblockA{\textit{dept. name of organization (of Aff.)} \\
% \textit{name of organization (of Aff.)}\\
% City, Country \\
% email address or ORCID}
% }

\maketitle

\begin{abstract}
As live streaming services skyrocket, Crowdsourced Cloud-edge service Platforms (CCPs) have surfaced as pivotal intermediaries catering to the mounting demand. Despite the role of stream scheduling to CCPs' Quality of Service (QoS) and throughput, conventional optimization strategies struggle to enhancing CCPs' revenue, primarily due to the intricate relationship between resource utilization and revenue. Additionally, the substantial scale of CCPs magnifies the difficulties of time-intensive scheduling. To tackle these challenges, we propose Seer, a proactive revenue-aware scheduling system for live streaming services in CCPs. The design of Seer is motivated by meticulous measurements of real-world CCPs environments, which allows us to achieve accurate revenue modeling and overcome three key obstacles that hinder the integration of prediction and optimal scheduling. Utilizing an innovative Pre-schedule-Execute-Re-schedule paradigm and flexible scheduling modes, Seer achieves efficient revenue-optimized scheduling in CCPs. Extensive evaluations demonstrate Seer's superiority over competitors in terms of revenue, utilization, and anomaly penalty mitigation, boosting CCPs revenue by 147\% and expediting scheduling $3.4 \times$ faster.

\end{abstract}

% \begin{IEEEkeywords}
% component, formatting, style, styling, insert
% \end{IEEEkeywords}

\section{Introduction}

The increasing popularity of live streaming services can be attributed to the widespread availability of high-speed networks (e.g., LTE/5G) and advanced personal devices (e.g., iPhone 14 Pro). As a result, live streaming has become one of the most prevalent applications in recent years (e.g., Twitch and TikTok). According to a Zippia report \cite{cicso}, live streaming services accounted for 17\% of global IP traffic in 2022 and are projected to grow at a rapid pace. In addition, live streaming has become an essential part of many internet users' daily routines, with the report \cite{cicso} showing that people spend 550 billion hours watching live streams in 2021.

\begin{figure}
    \centering
    \includegraphics[width=8cm]{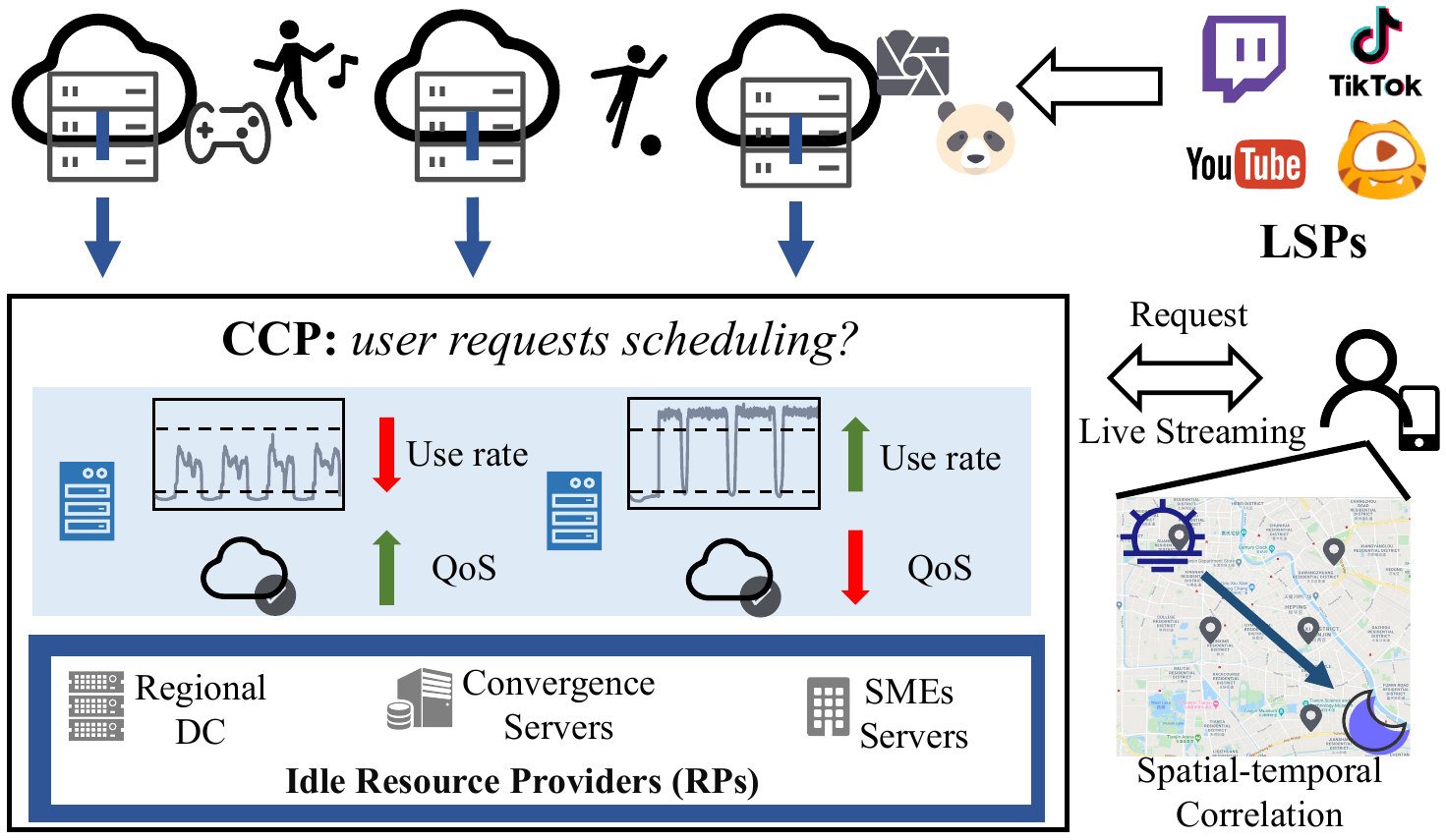}
    \caption{Crowdsourced Cloud-edge service Platform.}
    \vspace{-15pt}
    \label{fig:CCP}
\end{figure}

This surge in live streaming services has led to the emergence of Crowdsourced Cloud-edge service Platforms (CCPs) that cater to the increasing demands. As illustrated in Fig. \ref{fig:CCP}, the CCP services various live streaming platforms (LSPs) and integrate idle heterogeneous computing and bandwidth resources from resource providers (RPs). Owing to its unique crowdsourcing characteristics, CCPs can expand server coverage for LSPs at a lower cost while enhancing end-users' Quality of Experience (QoE) through closer proximity services \cite{xu2021cloud}. The ability to schedule live streaming requests is at the core of CCPs' Platform-as-a-Service (PAAS) offering for LSPs, acting as a proxy and managing a portion of LSPs' service capabilities. As such, scheduling plays a decisive role in CCPs' Quality of Service (QoS) and revenue \cite{9355041}.

% Existing geographic proximity-based scheduling strategies are overly simplistic and inadequate to address the complex challenges in optimizing cloud-edge platforms' utilization. Previous works have primarily focused on reducing cost and enhancing users' Quality of Experience (QoE), but have not adequately addressed the optimization of platform revenue, a crucial aspect of these platforms.

% In the context of existing geographic proximity-based scheduling strategies being overly simplistic, several research works have focused on reducing cost and enhancing user QoE. For example, works [a], [b], and [c] mainly concentrate on employing data-driven prediction strategies and optimizing viewer-side performance to improve user experience. However, these studies have not made significant advancements in optimizing the revenue of edge-cloud service platforms.

% These existing approaches tend to emphasize the last-mile (i.e., viewer-side) performance, such as works [6], [8], and [9], which rely heavily on viewer feedback and require quick responses, potentially limited in real-time scenarios. Consequently, these methods fall short in addressing the revenue optimization problem for edge-cloud service platforms.

In the context of existing geographic proximity-based scheduling strategies being overly simplistic, numerous studies have begun to employ data analysis and prediction of massive scheduling log data to develop optimization strategies or algorithms for achieving more optimal scheduling, such as frameworks aimed at maximizing user QoE or minimizing platform costs \cite{Li2022, Zhang2020, Haouari2019, zhang2022aggcast, jiang2016cfa, 10154347}. Although effective, these studies have offered minimal help in improving the revenue of CCPs. As a third-party matchmaking platform, CCPs consolidate a substantial amount of idle resources and strive to maximize their revenue by expanding the scale of services provided on these resources. Therefore, improving resource utilization is the key to increasing CCPs' revenue.

In fact, the resource utilization promotion brings benefits to all three parties involved: CCPs, LSPs, and RPs. For LSPs, higher utilization means that more requests can be processed at a lower cost \cite{zhang2022aggcast}. For crowdsourced RPs, higher utilization translates to increased income, thus boosting their engagement. As can be seen, a good resource utilization optimization schedule can promote the thriving development of the entire crowdsourced live streaming service ecosystem.

Nonetheless, tackling this issue is highly challenging due to the intermediary nature of CCPs. There exists a complex, non-linear relationship between server resource utilization and CCPs' revenue, which is significantly influenced by RPs' engagement and the QoS Service Level Agreements (SLA) from LSPs. Existing works fall short in addressing this issue and are ill-suited to the revenue optimization task.

Furthermore, as CCPs act as the scheduling proxy for LSPs, the time consumed in generating and executing scheduling strategies is crucial, an aspect often overlooked in prevailing research. Although scheduling models are becoming increasingly sophisticated, the user-perceived latency can significantly increase if the scheduling process takes an extended time. Given that CCPs need to handle numerous heterogeneous servers and diverse live streaming requests simultaneously, the time cost within such a vast scheduling space is formidable.

% (such as the issuing platform, content category, location, and quantity)
Anticipating the distribution of incoming requests is a potential method to enhance scheduling efficiency \cite{duc2019machine, 10.1145/3580305.3599453, 10.1145/3459637.3482298}. However, existing methods struggle to integrate request prediction with actual scheduling. This process mainly encounters three obstacles: 1) The inability to simultaneously predict multiple heterogeneous characteristics of requests. 2) Difficulty in foreseeing the revenue of deploying requests on heterogeneous servers. 3) The challenge of dealing with the discrepancies between predicted and actual request distributions. Additionally, the real-world CCP environment's intense spatiotemporal fluctuations complicate request prediction itself.

% 1) The inability to simultaneously predict multiple heterogeneous characteristics of requests that affect scheduling.
% Moreover, in a real-world CCP environment, the intense spatialtemporal fluctuation of user requests pose challenges to the prediction of requests themselves.

% 旧版：incorporating revenue into the scheduling optimization introduces an extra dimension, shifting from a one-dimensional approach (solely focusing on service cost) to a two-dimensional perspective (considering both cost and revenue concurrently). 

% 由于CCP的中间平台属性，服务器的利用率与CCP收益之间存在着复杂的非线性关系。现有的工作，不论是将资源利用率视为服务成本(站在LSP角度)，还是收益(站在资源提供商的角度)，都不能完善妥当的处理这一问题。

% 此外，CCP作为LSP的调度代理，产生和执行调度策略所消耗的时间也至关重要，这一问题在现有的研究中多被忽略。尽管调度模型被设计的愈加精细，但如果调度过程占用了较长的时间，用户在观看直播时感知到的时延也会明显增加。由于CCP需要同时处理大量的异构服务器和多样的LSP直播请求，实时的调度方法在如此庞大的调度空间中所产生的时间成本是用户无法接受的。

% 提前预测未到的请求分布是提高调度效率的有效潜在方法\cite{}，然而现有的方法都没能有效地将请求预测与真实的调度结合起来，这一过程主要面临三个阻碍：1）无法同时预测请求的多种异构特征（例如发出平台，内容类别，位置和数量），而这些特征都对调度有着较大影响。2）难以预知请求在异构服务器上部署后的效果。 3）难以处理预测与真实到来请求分布之间的误差，而保证所有真实请求被成功服务是一个调度算法现实可用的基本条件。

% 此外，在现实的 CCP 环境中，用户请求强烈的时空波动特性给请求预测本身也带来了挑战。

% Furthermore, in realistic CCP settings, users' requests exhibit strong spatialtemporal fluctuations, making optimization strategies subject to the complex feature variations caused by heterogeneous requests.

% 进一步的，我们提出了Seer: an efficient revenue-aware live-streaming request scheduling system within the CCP. 它成功地将请求预测与调度优化相结合并通过创新的预调度-重调度模式实现了高效的收益最优化调度，同时保证了CCP的收益最大化，调度效率的最大化，并提供了灵活的配置选项。

To tackle these challenges, we propose a comprehensive solution that focuses on refining the scheduling process to augment CCP revenue. Specifically, we employ a large-scale dataset of real-world live streaming services in CCP to model the intricate relationship between resource utilization and CCP revenue. Furthermore, we introduce Seer, a proactive revenue-aware live streaming scheduling system. Seer successfully integrates request prediction with scheduling optimization, leveraging an innovative \textbf{Pre-schedule-Execute-Re-schedule (PER)} paradigm to attain efficient revenue optimization. Simultaneously, it ensures maximal revenue for CCP, and scheduling efficiency, whilst offering flexible configuration options. Our contributions are summarized as follows:

% This groundbreaking approach serves to demonstrate how significant improvements in resource utilization, service quality, and revenue can be achieved within the context of real-world CCP platforms.

% In addition, we investigate the revenue and QoS cost implications of requests with varying characteristics. Our prediction model, based on spatial-temporal features for each request category, allows for more accurate predictions under real-world constraints. By combining our data-driven insights with optimization techniques, we present a novel approach to address the challenges in revenue optimization for CCP effectively.

% In this paper, we tackle the challenge of improving platform revenue (device utilization) while ensuring QoE and Quality of Service (QoS) by refining the scheduling process. Our work overcomes the challenges posed by requests with both revenue attributes and service costs, fluctuating spatialtemporal distributions of requests, and the combined impact of request types and server QoS on user QoE.

% To achieve this, we first perform data analysis to classify requests. Next, we quantify the impact of each request category on server QoS and revenue, as well as the effects of request categories and server QoS on user QoE. Then, we propose a prediction model based on spatialtemporal features for each request category. Building upon the prediction results, we model the platform's request scheduling problem and, using optimization algorithms, maximize device utilization under realistic constraints while ensuring server QoS and user QoE.

\begin{itemize}[leftmargin=2em]
\item[($i$)] To the best of our knowledge, we are the first to focus on the revenue optimization of CCPs serving the emerging live streaming industry.
\item[($ii$)] Utilizing a comprehensive dataset from real-world CCP operations, we devise a nuanced model that delineates the relationship between resource utilization and CCP revenue. This process also involves an in-depth analysis of requests and allows us to identify and isolate critical features that impact scheduling issues.
\item[($iii$)] We introduce Seer, a proactive revenue-aware scheduling system that combines request prediction and scheduling optimization through a novel Pre-schedule-Execute-Re-schedule paradigm. Evaluations based on real-world data demonstrate that Seer significantly enhances CCPs revenue and reduces time costs.
\end{itemize}

% The rest of this paper is organized as follows: In Section \ref{rw}, we briefly introduce the related works. A formal definition of the prediction problem is given in Section \ref{npd}. Section \ref{modeling} introduces the user-centric modeling methods. Section \ref{framework} presents the proposed T-EEGCN in details. Section \ref{Exp} shows the experiments results of the proposed model and the ablation studies. We finally conclude this paper in Section \ref{conclusion}.

\section{related work}\label{rw}
% In this section, we review previous researches regarding live streaming service architectures (including LSP and CCP) and their optimization strategies.

\textbf{LSPs and Crowdsourced Cloud-edge Architectures.}  The rapid expansion of live streaming services can be attributed to the technological advancements in multimedia delivery services \cite{giuliano2020integration, jedari2020video}. From a technological standpoint, LSPs encompass a video delivery service that records and broadcasts media content to all users in real time. The coexistence of simultaneous content generation and user playback presents substantial challenges for server transmission quality \cite{10.1145/3519552}.

% For LSP, the process of generating and delivering personal live streaming content to end-users involves several servers and transmission links. The simultaneous occurrence of media content generation and user playback behavior poses significant challenges for server communication bandwidth and service quality.

To address these challenges, LSPs have started employing CCPs to reduce content distribution costs and get lower delivery latency \cite{adhikari2012unreeling, zhang2019enhancing}. By integrating widespread idle computing and transmission resources from RPs, CCPs offer a promising solution for LSPs. Compared to providing standardized servers \cite{MANVI2014424}, it is more common for CCPs to serve as an independent platform handling requests for LSPs (PaaS) \cite{pahl2015containerization}. In a PaaS setup, LSPs directly forward requests to CCPs, which then select the most optimal server. In this regard, CCPs can be viewed as service proxies for LSPs\cite{zhang2022aggcast}.

% , performing similar functions 

Analogous to LSPs, CCPs face the challenge of handling massive requests with limited resources. The situation is even more demanding for CCPs, as they must guarantee the LSPs' QoS SLA, while ensuring the engagement of RPs \cite{zhu2021user, anglano2018profit}.

\textbf{Optimization Strategies for Live Streaming Service Platforms} have garnered considerable attention from both the research community and industry since they provides more significant improvements at lower costs than modifying transfer protocols and architectures \cite{barakabitze2019qoe}. Current industry scheduling mainly rely on geography-based assignment, neglecting request characteristics and factors like QoS and QoE \cite{dilley2002globally}.

To address these issues, some studies have proposed improved scheduling strategies. Zhu et al.\cite{zhu2021user} utilizes decision trees to capture user video engagement, mapping requests to maximize overall QoE. Zhang et al.\cite{8485962} implements peer-to-peer transmission and proactive high-demand content distribution to decrease server bandwidth usage. Zhang et al.\cite{zhang2022aggcast} suggests aggregating dispersed audiences and assigning them to fewer pre-selected nodes to reduce bandwidth costs. More studies focus on guaranteeing QoE while lowering service costs. Some suggest adaptive content uploading and edge prefetching strategies \cite{Li2022, Haouari2019}, while others propose deep reinforcement learning-based (DRL) edge-assisted multicast frameworks for real-time request decisions \cite{8737456, zhang2020leveraging, 9047133, 9155467}.

% To address these issues, some studies have proposed improved scheduling strategies. Zhu et al.\cite{zhu2021user} utilizes decision trees to capture user video engagement, mapping requests to maximize overall QoE. Other studies focuses on reducing content distribution costs. Zhang et al.\cite{8485962} implements peer-to-peer transmission and proactive high-demand content distribution to decrease server bandwidth usage. Zhang et al.\cite{zhang2022aggcast} suggests aggregating dispersed audiences and assigning them to fewer pre-selected nodes to reduce bandwidth costs. More studies focus on guaranteeing QoE while lowering service costs. Some suggest adaptive content uploading and edge prefetching strategies \cite{Li2022, Haouari2019}, while others propose deep reinforcement learning-based (DRL) edge-assisted multicast frameworks for real-time request decisions \cite{8737456, zhang2020leveraging, 9047133, 9155467}.

Although effective, these methods fall short when applied to the revenue-optimized scheduling of CCPs due to two factors. 1) The intermediary nature of CCP necessitates a simultaneous consideration of LSPs' QoS SLA and RPs' engagement, leading to a complex nonlinear relationship between resource utilization and revenue. 2) CCPs are confronted with a vast number of servers and diverse requests. Current real-time scheduling strategies (e.g., the DRL-based frameworks) may introduce scheduling latency that is formidable.

% Although effective, these methods 都无法很好地用于CCP的收益优化调度中，正如前面所说，这是由于1）CCP的中间人特性导致其需要同时考虑LSP的QoS SLA和RP的参与，使得其利用率与收益之间存在复杂的非线性关系（不能一味最小化或最大化利用率）。2）CCP面临庞大的服务器数量和多样的请求，现有的实时调度策略（如基于DRL的框架）可能会引入难以接受的调度时延。

% overlook the revenue aspect of service platforms, which requires addressing trade-offs between service costs and revenue for requests. Additionally, we regard that existing request prediction research often relies on assumptions of compliance with historical distributions, without considering spatialtemporal fluctuations, which indirectly affects the performance of optimization models in real-world environments. Finally, due to CCP's intermediary nature, revenue optimization must address dual constraints of SLAs for user QoE and LSP for CCP service QoS.

\section{MEASUREMENTS AND MOTIVATIONS}\label{npd}
In this section, we shed light into the following questions:
\begin{itemize}[leftmargin=*]
    \item Describe the dataset, QoS metrics and request revenue.
    \item Validate potential improvement and optimal revenue range.
    \item Examine the spatialtemporal dependencies of requests.
    \item Investigate the features affecting  the request revenue.
\end{itemize}
By addressing these questions, we aim to gain insights that enable the modeling of the link between resource utilization and CCPs revenue, and facilitate the construction of a robust scheduling framework.

\subsection{Dataset Description and Features Describing}
Our research is conducted in collaboration with a leading CCP in China, which consists of 5174 servers distributed across the country and serves more than five typical LSPs. With the CCP's assistance, we collected three types of data:
\begin{itemize}[leftmargin=2em]
\item[($i$)] \textbf{Live streaming service logs (LL):} We collected 10 days of service logs from all servers dedicated to live streaming. This dataset contains 17854 live streaming channels, 476 unique server IDs (which can be linked to server attributes, such as bandwidth and location), and 500 million requests (with a specific channel and bitrate) across 59 locations. For each unique session, the server records the data size transmitted in one-minute intervals.
% We collected 10 days of service logs from all servers dedicated to live streaming, amounting to 500+ GB. This dataset contains 17854 live streaming channels, 476 unique server IDs (which can be linked to server attributes, such as bandwidth and location), and 500 million requests (with a specific channel and bitrate) across 59 locations. For each unique session, the server records the data size transmitted in one-minute intervals.
\item[($ii$)] \textbf{Client-reported logs:} These logs correspond in time span to the LL dataset. Every 60 seconds, clients upload the total number of requests issued by users and the related metrics, of which we focus on the startup latency as it is the performance metric that best characterizes the QoE.
% \item[($iii$)] \textbf{Server-side QoS logs (SL):} Similar to the client-reported QoE logs, every 60 seconds, the server records the total number of received requests, the successfully served requests, and the failed requests due to cache misses, server overloads, resulting in service stalling, timeouts, or even disconnections (we calculate the proportion of all types of failed requests as the server's QoS).
\item[($iii$)] \textbf{Server-side logs:} Similar to the client-reported logs, every 60 seconds, the server records all the successfully served requests, and the failed requests due to server overloads timeouts, or even disconnections (we calculate the error rates of all types of failed requests).
\end{itemize}

For the sake of convenience, we refer to the startup latency and requests error rates as the server's QoS, given that they both form the basis for LSPs' evaluation of the CCP's QoS SLA. Furthermore, certain studies consider startup delay and related stall metrics as QoS indicators \cite{zhu2021user, zhang2022aggcast}. Besides, we also refine the notion of \emph{request-server revenue}, denoting \emph{the average data throughput per minute for a request on a given server}, herein termed as \textbf{request revenue}.
% 除了定义CCP的整体收益，为了更好地建模，我们将一个请求在某服务器上被服务时每分钟所产生的平均数据传输size称为请求-服务器收益，简称请求收益。
% we consider the data size requested per minute as the revenue for a request. 

\textbf{Ethical Considerations}: We have taken a series of measures to ensure the ethical use of data. All user information, including user IDs, IP addresses, and even live streaming room IDs, is anonymized. We do not link the service logs to users; instead, we analyze the service logs at the metadata level (such as the request time and content category).

\subsection{Potential Improvement and Optimal Revenue Range}\label{3b}
\begin{figure}
\centering
\subfigure[CDF of server utilization]{
\includegraphics[width=4.2cm]{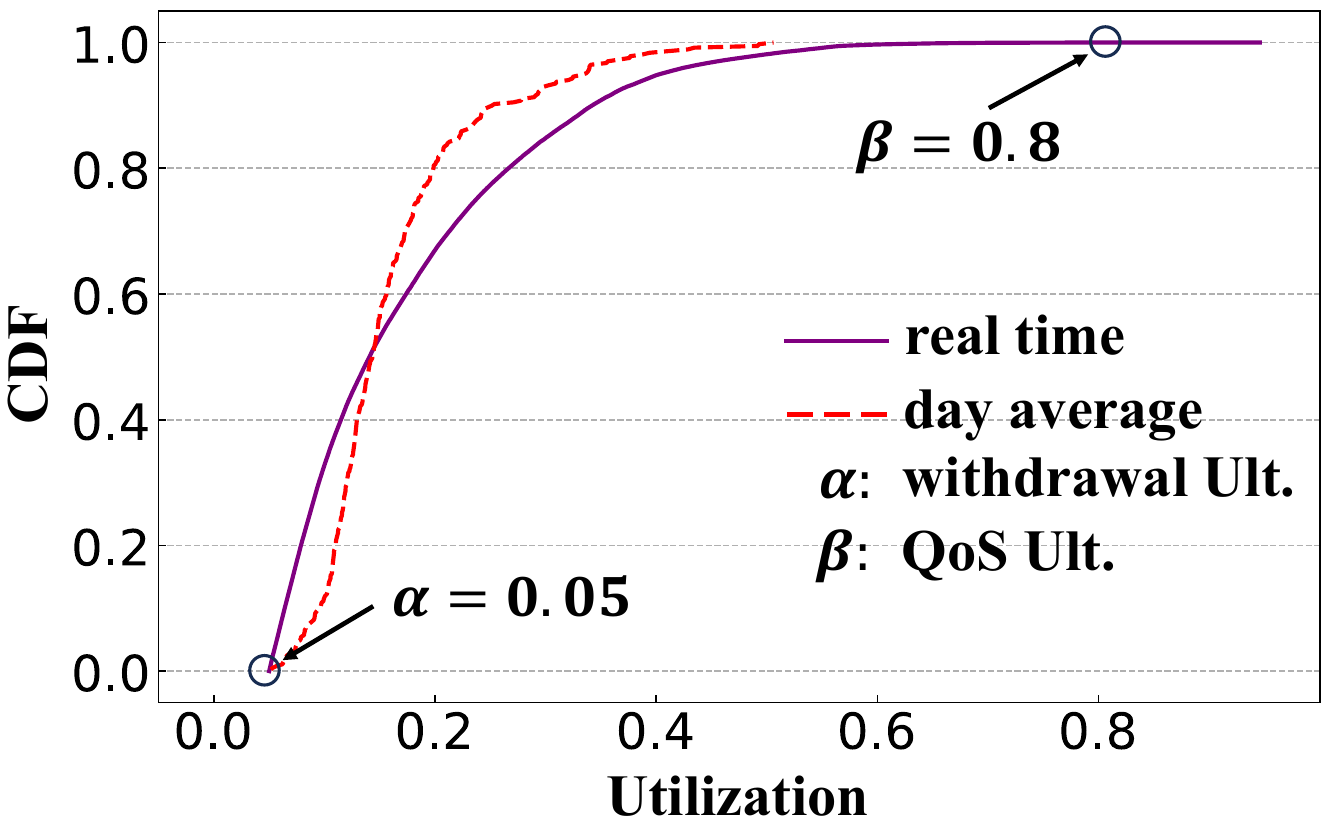}\label{fig:cdf_ut}
}
\hspace{-2mm}
\subfigure[24-hour workload]{
\includegraphics[width=3.5cm]{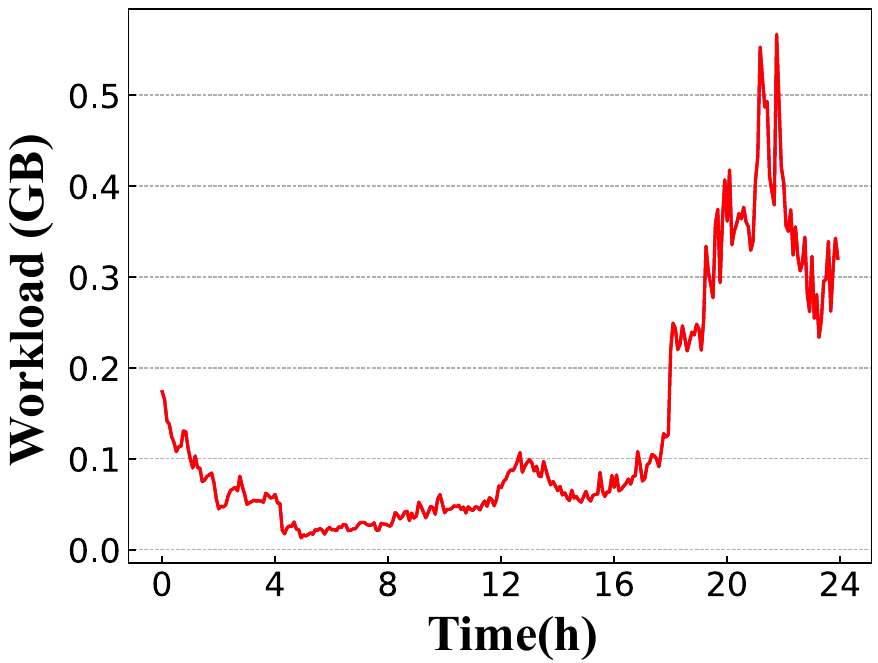}\label{fig:ut_time}
}
\caption{Server utilization analysis.}
\vspace{-15pt}
\end{figure}

% 思考是否将潜在提升部分放在Intro部分，以对比图说明。
We first investigate the server resource\footnote{Given the nature of live streaming services, all references to 'resources' in the following sections specifically denote bandwidth.} utilization in a typical day. Fig. \ref{fig:cdf_ut} presents the cumulative distribution function (CDF) of the utilization of all edge servers. As the figure illustrates, a majority of servers experience remarkably low utilization; specifically, more than 60\% of servers exhibit real-time utilization of less than 20\% throughout the day.

Such low utilization is not uncommon for CCPs. Fig. \ref{fig:ut_time} shows that the workload volume tend to cluster within two relatively short periods - around noon and evening, while only a handful of requests are observed during the remainder of the day, leading not only to resource wastage but also additional power and operational costs. We argue that such challenges can be alleviated through intelligent recognition of request and scheduling geared towards optimized utilization.

% Such low utilization is not uncommon for CCP due to inherent conflict between the wide distribution of servers, the prolonged operational periods, and the characteristic patterns of live streaming services. Fig. \ref{fig:ut_time} shows that the workload volume tend to cluster within two relatively short periods - around noon and evening, while only a handful of requests are observed during the remainder of the day.

% indiscriminately
% The conventional geographically-proximate scheduling algorithms ignore this pattern, assigning requests to the closest server. This strategy results in a substantial portion of servers experiencing long periods of low utilization, leading not only to resource wastage but also additional power and operational costs. We argue that such challenges can be alleviated through intelligent recognition of request distribution and optimization of scheduling geared towards optimized utilization.

\begin{figure}[hb]
    \centering
    \includegraphics[width=5cm]{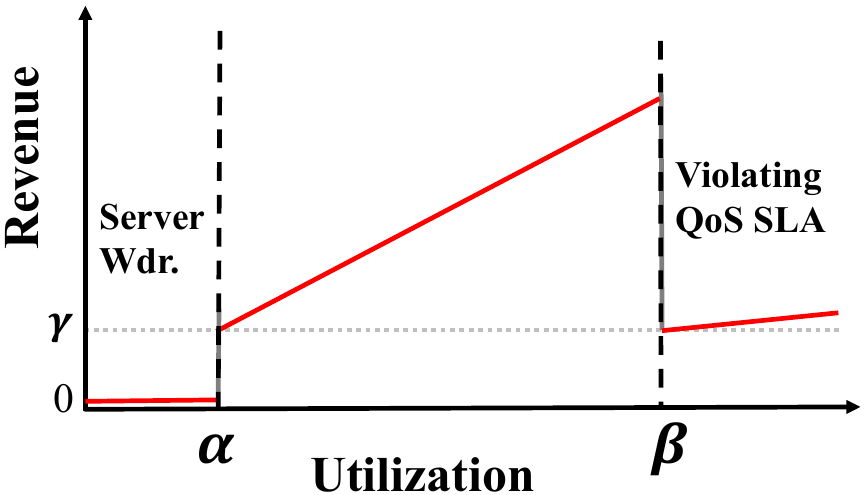}
    \caption{Fitting plot of server utilization versus revenue.}
    \vspace{-5pt}
    \label{fig:ut_rev}
\end{figure}

Besides the potential improvement, we can observe from Fig. \ref{fig:cdf_ut} that server utilization ranges from a lower bound $\alpha$ to an upper bound $\beta$ (with $\alpha = 0.05$ and $\beta = 0.8$). The presence of the lower threshold can be attributed to the fact that most of CCP's servers are leased from RPs, who employ the resource utilization as a measure of their revenue. To guarantee a minimum income, RPs set a minimum utilization threshold. If CCP's scheduling results in server utilization frequently falling below the threshold, RPs may terminate their agreement with CCP and withdraw servers. We refer to this threshold as the server \textbf{withdrawal utilization $\alpha$}.

Due to limitations in network conditions, server hardware and other factors, resource utilization can never reach infinity. As server load increases, the QoS tends to degrade. In practice, CCP signs QoS SLA with LSPs to ensure a minimum QoS. This SLA, reflected in terms of utilization, sets an upper utilization threshold $\beta$, which we call the \textbf{QoS utilization}.

With the two utilization thresholds, both closely related to CCP's revenue, we can clearly model the relationship between server resource utilization and CCP's revenue. From Fig. \ref{fig:ut_rev}, it is evident that when utilization falls below $\alpha$, the associated revenue could be negligible due to the risk of server withdrawal\footnote{Although servers aren't instantly withdrawn from CCP, the maintenance cost equals or surpasses the server's gain, resulting in negligible revenue.}. Once server utilization exceeds $\alpha$, a linearly positive relationship can be established. This implies that CCP's revenue increases proportionately with the utilization. Finally, upon reaching the QoS utilization $\beta$, CCP's revenue inevitably diminishes to a lower value $\gamma$ due to the substantial penalties incurred from violating the QoS SLA. The exact penalty imposed depends on the agreement between CCPs and LSPs. For the sake of modeling, we set $\gamma$ to fluctuate between 10-30\% of the original revenue to represent the penalties LSPs impose on CCPs for SLA violations\cite{10.1145/2843890}.

In the scheduling process, the CCP aims to stay within a profitable utilization range (optimal revenue range). The withdrawal utilization can be managed by balancing server resources and assigned requests. However, handling the QoS utilization is more complex. This threshold acts as utilization feedback when a QoS SLA is violated, and can't be sufficiently controlled merely by capping tallied utilization. Therefore, it is crucial to establish a mapping from QoS to $\beta$ to guide the scheduling process (detailed in Section \ref{prescheduling}).

\subsection{Spatial-Temporal Correlation of Requests}
In this part we explore the dependency of requests in both the temporal and spatial domains, thereby offering guidance for subsequent request prediction tasks.

\begin{figure}
\centering
\subfigure[Fluctuation of requests]{
\includegraphics[width=4.0cm]{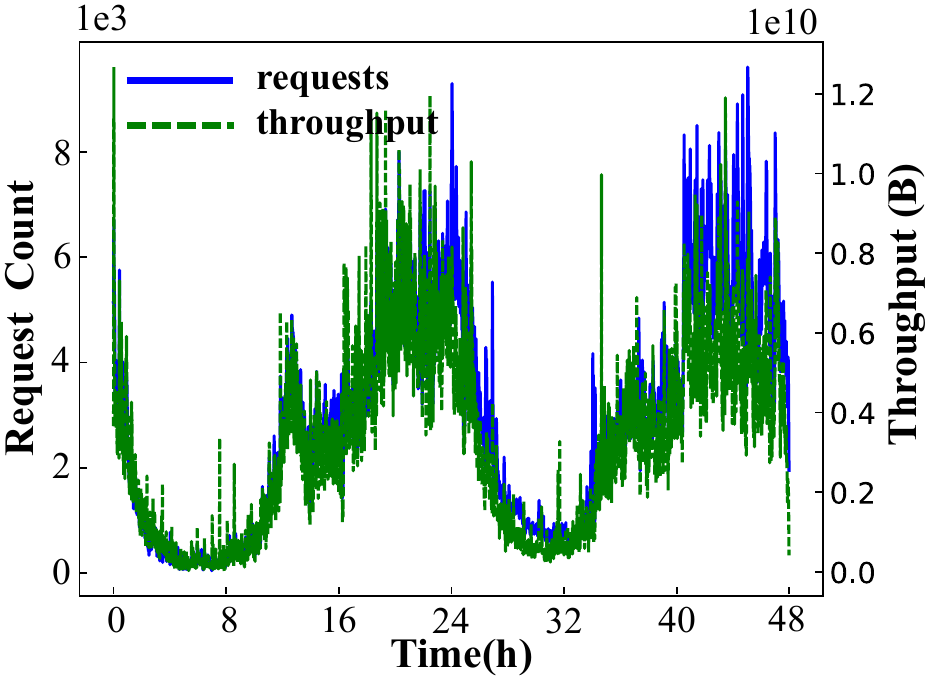}\label{fig:temporal}
}
\hspace{-3mm}
\subfigure[Temporal autocorrelation]{
\includegraphics[width=4.2cm]{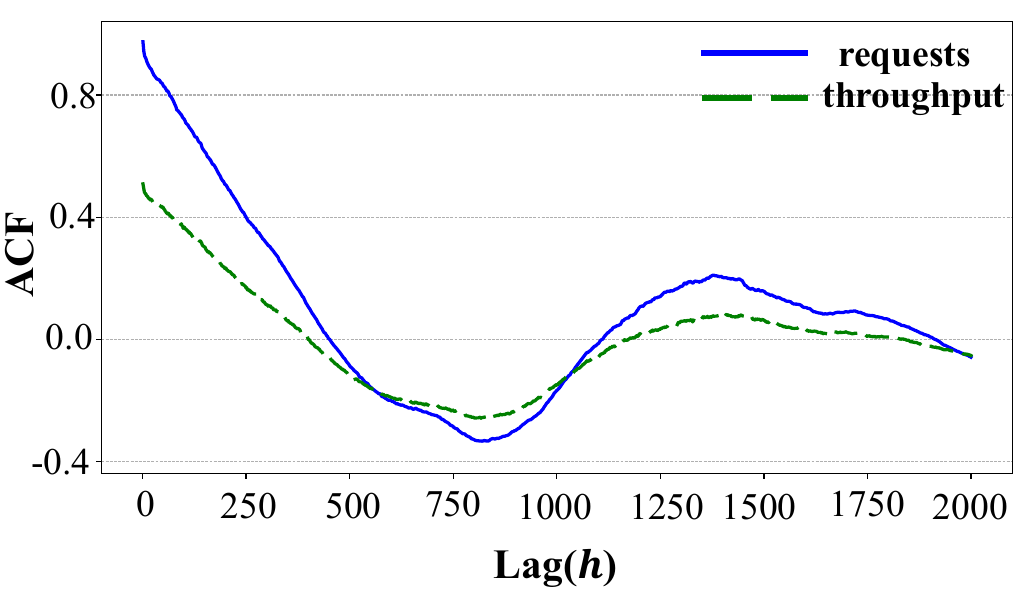}\label{fig:temporal-acf}
}
\caption{Temporal fluctuation and autocorelation analysis.}
\vspace{-8pt}
\end{figure}

% That is, the request volume is generally similar to the historical volume during the corresponding time period. 
From a temporal perspective (e.g., Fig. \ref{fig:temporal}), requests show periodic variations that align with the day's progression. Moreover, due to the gradual shifts in users' viewing interests and the infrequency of sudden request surges, future requests generally follow a gradual trend based on recent history. 

 % (which can be viewed as a sequence of samples representing all requests within a time period) 

To validate these observations, we employ the \emph{sample AutoCorrelation Function} (ACF) to investigate the temporal dependencies of requests. The ACF quantifies the dependency between values in a sampled process as a function of the time lag $h$. The ACF calculation procedure for a typical server can be formalized as follows:
\begin{equation}
\rho(h)=\frac{\sum_{t=1}^{D-|h|}\left(d_{t+|h|}-\bar{d}\right)\left(d_t-\bar{d}\right)}{\sum_{t=1}^D\left(d_t-\bar{d}\right)^2},-D<h<D
\end{equation}where $d_t$ is the request volume of time $t$, and $D$ and $\bar{d}$ are the total count and mean value of sampled requests in the temporal dimension, respectively. The autocorrelation value lies in the range $[-1, 1]$. $\rho(h) = 1$ indicates total positive autocorrelation between data with a time lag of $h$; while $\rho(h) = -1$ means total negative autocorrelation.

Fig. \ref{fig:temporal-acf} shows the sample ACF at time lag $h = 0, 1,..., 2000$ (unit in minutes) for both requests and throughput. From Fig. \ref{fig:temporal-acf}, we can see that the autocorrelation gradually decreases as $h$ increases, i.e., the future request volume depends mainly on the data from the most recent historical time. Besides, when the time lag equals 1380 (about one day), the autocorrelation is relatively high. This shows that the requests follows a clear daily pattern. For example, the requests peak and off-peak hours are similar on each day.
\begin{figure}
\centering
\subfigure[Spatial correlation of requests]{
\includegraphics[width=4.1cm]{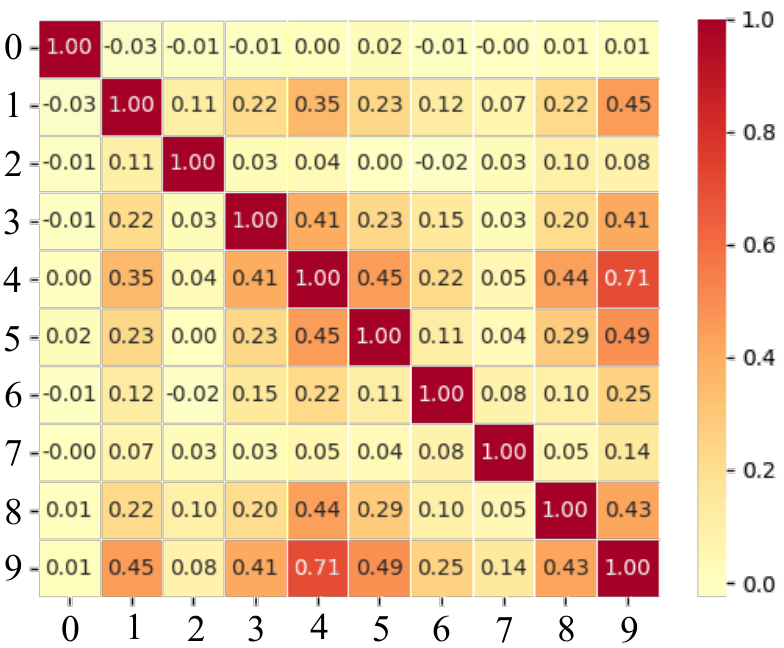}\label{spatial_req}
}
\hspace{-3mm}
\subfigure[Spatial correlation of throughput]{
\includegraphics[width=4.1cm]{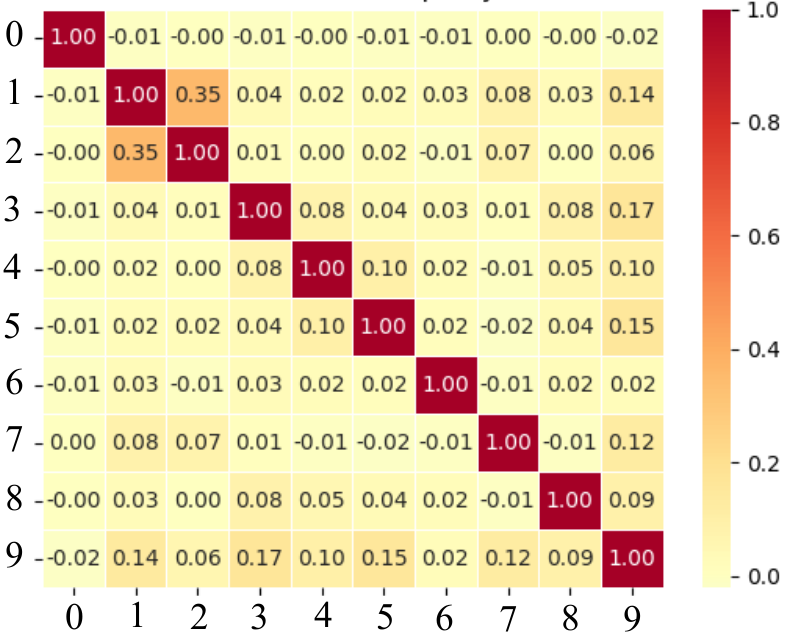}\label{spatial_data}
}
\caption{Spatial correlation analysis.}
\label{fig:spatial}
\vspace{-10pt}
\end{figure}

On the other hand, we examine the requests correlation in the spatial domain by calculating the Pearson correlation coefficient between two locations:
\begin{equation}
\rho=\frac{\operatorname{cov}\left(\mathbf{d}_{i}, \mathbf{d}_{j}\right)}{\sigma_{\mathbf{d}_{i}} \sigma_{\mathbf{d}_{j}}}
\end{equation}where $\operatorname{cov}$ is the covariance operator, $d_i$ is the request volumes at location $i$, and $\sigma$ is the standard deviation. The magnitude of the Pearson correlation coefficient is the same as ACF.

Fig \ref{fig:spatial} illustrates the spatial correlation of requests in terms of both the number of requests and data throughput. Specifically, Fig. \ref{spatial_req} shows a strong positive correlation between the number of requests across different locations, with the correlation coefficient reaching up to 0.71 in some instances.

The preceding analyses confirm the distinct spatialtemporal dependency of requests, prompting us to devise a spatial-temporal prediction model to enhance scheduling efficiency. However, Figures \ref{fig:temporal-acf} and \ref{spatial_data} indicate that, unlike request volume, data throughput doesn't display similar spatialtemporal patterns as throughput generated by different requests (i.e., request revenue) varies significantly. By investigating features that affects this variation in request revenue, we can obtain a method to foresee request revenue, a crucial factor in combining prediction and scheduling effectively.

% 这里加一些验证的话，证明请求可以通过时空特征提取来预测
% 图五从请求数量和数据吞吐量两个维度展示了请求的空间相关性，从图5a可以看出，不同地区之间的请求数量呈现较为明显的正相关（相关性系数最高能达到0.71）。

% 以上的分析证明了请求在较为明显的时空依赖性，这启发我们设计一个轻量化的时空预测模型来预测请求的分布，并通过预测能力提高调度效率。另外，从图4b和图5b可以看出，对比请求数量，数据吞吐量分布并没有呈现出非常明显的时空依赖性，这说明服务不同请求所产生的吞吐量（即请求收益）还受到请求自身特征的影响，我们可以通过调查这些影响特征来获得预知请求收益的方法，而解决这一问题是将预测与调度相结合的关键。

% \begin{figure*}[ht]
% % \setlength{\belowcaptionskip}{-0.5cm}
% \centering
% \subfigure[Start-up latency of different request platforms]{\label{fig:latency_platform}
% \includegraphics[height=2.5cm, width=4.8cm]{Pics/latency_platform.pdf}
% }
% \subfigure[error rate of different request platforms]{\label{fig:error_platform}
% \includegraphics[height=2.5cm, width=4.8cm]{Pics/error_platform.pdf}
% }
% \subfigure[Data transfer rate of different request platforms]{\label{fig:Mbpm_platform}
% \includegraphics[height=2.5cm, width=4.8cm]{Pics/Mbpm_platform.pdf}
% }
% \hspace{5mm}
% \subfigure[Startup latency for same-region and cross-region assignment]{\label{fig:latency_cross_region}
% \includegraphics[height=2.5cm, width=4.8cm]{Pics/latency_cross_region.pdf}
% }
% \subfigure[Start-up latency in different periods]{\label{fig:latency_period}
% \includegraphics[height=2.5cm, width=4.8cm]{Pics/latency_period.pdf}
% }
% \subfigure[error rate in different periods]{\label{fig:error_period}
% \includegraphics[height=2.5cm, width=4.8cm]{Pics/error_period.pdf}
% }
% \caption{Impact of request features on QoS.}
% \end{figure*}

\begin{figure}
\centering
\subfigure[Content category]{\label{fig:content}
\includegraphics[width=4.1cm]{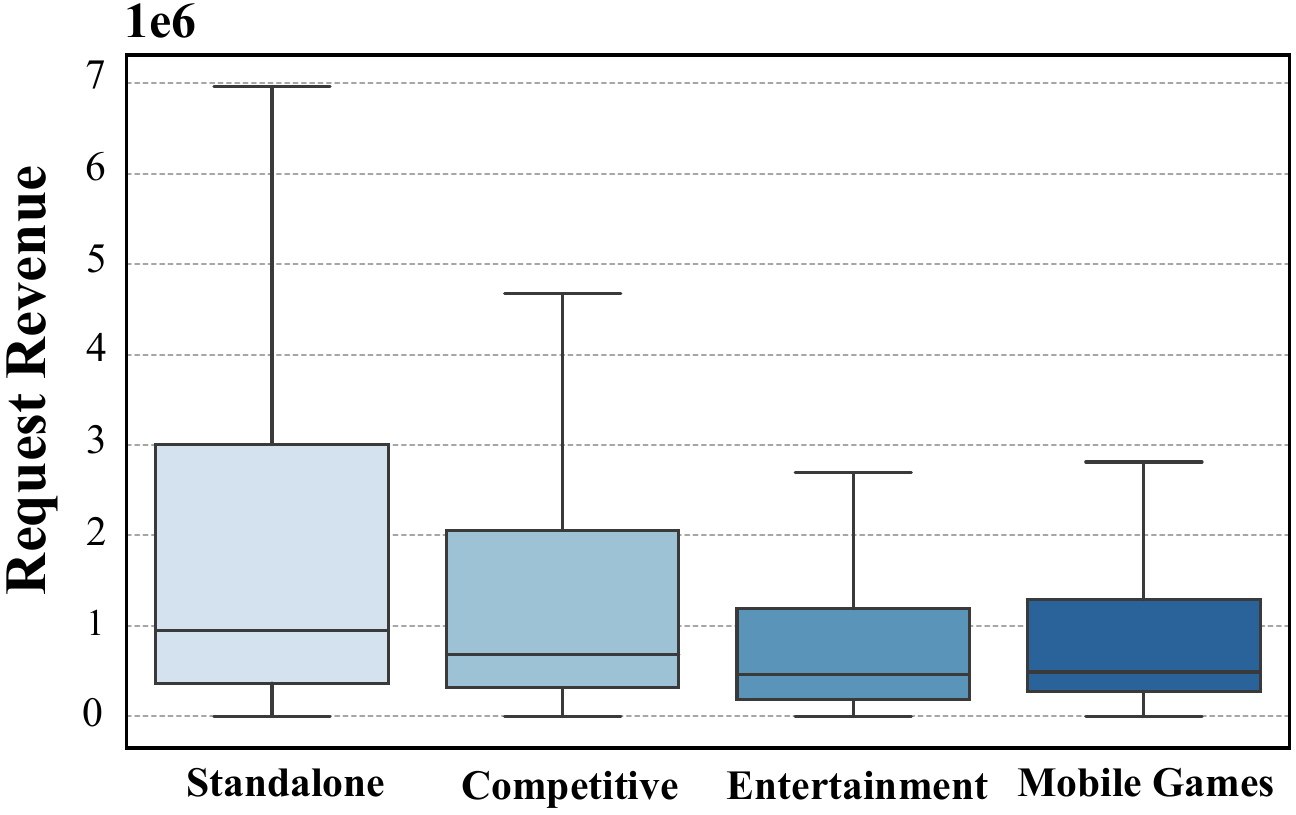}
}
\hspace{-3mm}
\subfigure[Platform]{\label{fig:platform}
\includegraphics[width=4.1cm]{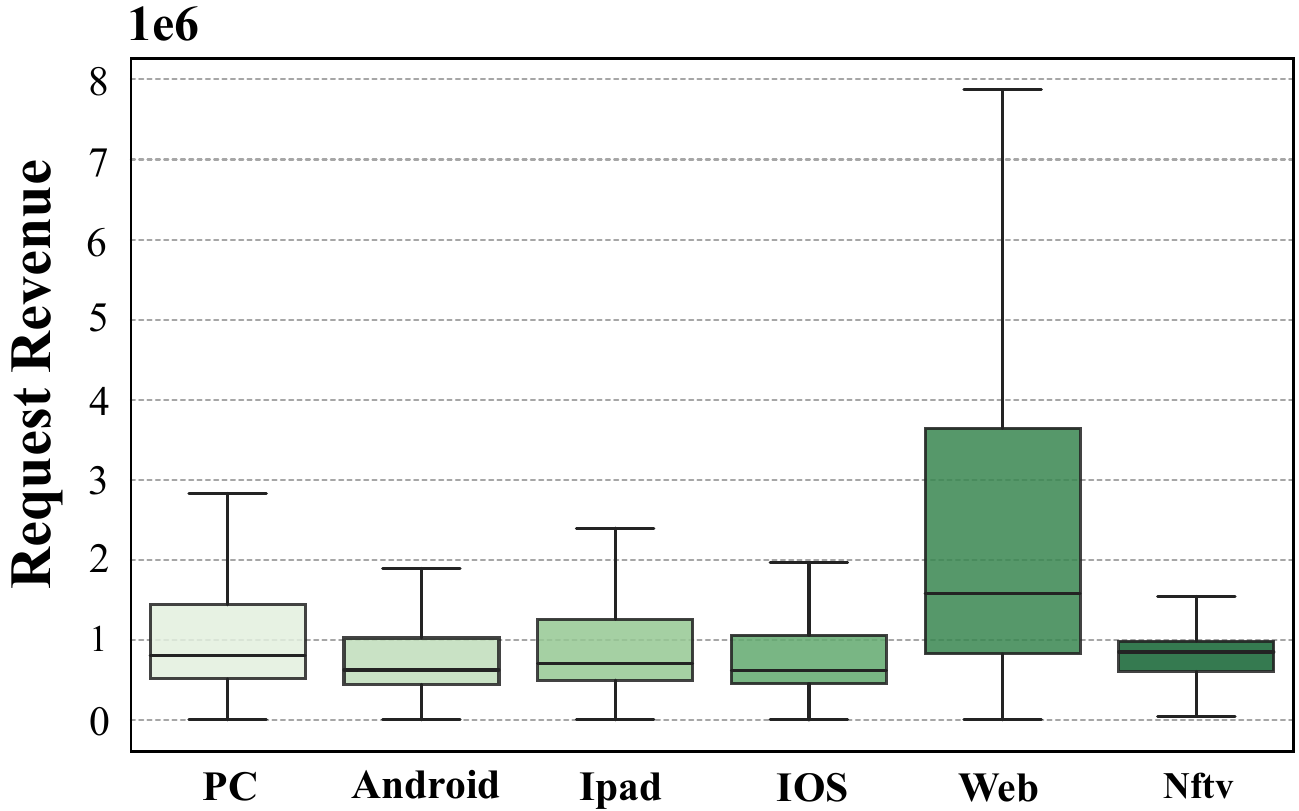}
}
\caption{Influence of different features on request revenue.}
\label{fig:features-rev}
\vspace{-10pt}
\end{figure}

\subsection{Principal Factors Affecting Request Revenue} \label{rev_ana} % 
% 为了将请求预测丝滑地运用到后续的收益优化调度中，我们需要解决三个关键的问题，即前文所提到的1)感知除时空分布以外的请求特征。2)foreseeing the revenue of deploying requests on heterogeneous server.3)Dealing with the discrepancies between predicted and actual request distributions.

% 在本部分我们重点解决前两个问题，

% 为减少调度过程的时间成本，我们建议提前遍历所有可能的请求-服务器组合并获得相应的部署收益，

We identify three key features that have the most significant impact on request revenue: a) content category, b) request platform, and c) request timing (peak or off-peak).

%首先我们依照请求所访问的直播内容对请求进行内容分类，我们将LL数据集中的所有请求记录划分为x个类别，如表x所示。图x(a)和图x(b)分别展示了不同内容类别的请求的平均收益和QoE(请求失败次数)占比，可以看出，网游竞技等类别的内容由于较高的码率和时延要求，会导致更高的收益和QoE成本，而诸如娱乐天地，手游等休闲类的直播内容对QoE的影响更小，但同时收益也会更低。进一步的，我们绘制了主要处理网游竞技类请求和休闲类请求的服务器之间的QoS对比图，如图x所示，处理网游竞技类请求的服务器往往拥有更高的错误请求数，反之处理休闲类请求的服务器往往能更轻松地完成服务，这也说明了请求的内容对服务器的QoS有着深刻的影响。为了让整个优化系统能更好地处理内容类别带来的多方面影响，我们将内容类别作为对请求进行聚类的关键特征之一，并在对系统的收益进行优化时限制每台服务器所处理的高需求类别的请求占比（后面加到insights里）。
Firstly, we classify all requests into four categories based on the live stream channel, as represented in Fig. \ref{fig:content}, which displays the varying revenue distribution across categories. Notably, requests for competitive gaming content yield higher revenue due to increased bitrate and latency requirements, whereas casual content such as entertainment and mobile game streams generate less revenue.

% 另一个请求收益的决定性特征是平台，即用户观看直播所使用的设备，这是由于请求平台往往影响着用户所观看的内容和请求的码率。从图6b可以看出，从更稳定的平台（例如PC上客户端和网页端）所发出的请求相比于移动平台（例如手机和ipad）具有更高的收益，即用户更倾向于在PC上选择更高清的码率，另外PC本身较好的网络条件也支持更高的数据吞吐量。

% 最后，我们选择request tming （即请求发出的时间是否属于一天的高峰期）作为最后考虑的特征，这一特征对请求收益的影响不如前者那么易于理解，但得益于合理的数据分析我们可以清晰的发现它的决定性作用。如图7所示，我们可以看到当服务器所处理的请求数量处于峰值时，每条请求的收益往往最低，而在请求相对稀少的时刻，请求的收益将会大大增加。我们认为这归因于在请求高峰时刻，用户所选择观看的视频码率质量相对平时反而有所下降，另外由于CCP服务压力较大，网络状况波动，每条请求产生的数据吞吐量也会小于平时。

Another determinant feature is the platform (i.e., the device used for viewing the live stream) since it influences both content type watched and the requested bitrate. As shown in Fig. \ref{fig:platform}, requests emanating from more stable platforms (e.g., PC and Web clients) yield higher revenues compared to mobile platforms (e.g., smartphones and tablets), indicating a preference for higher quality streams on PC and the inherently better network conditions supporting higher data throughput.

% While the impact of this feature on request revenue isn't as intuitive, thorough data analysis reveals its decisive role. 

\begin{figure}[ht]
    \centering
    \begin{minipage}{0.5\columnwidth}
        \centering
        \includegraphics[width=4.2cm]{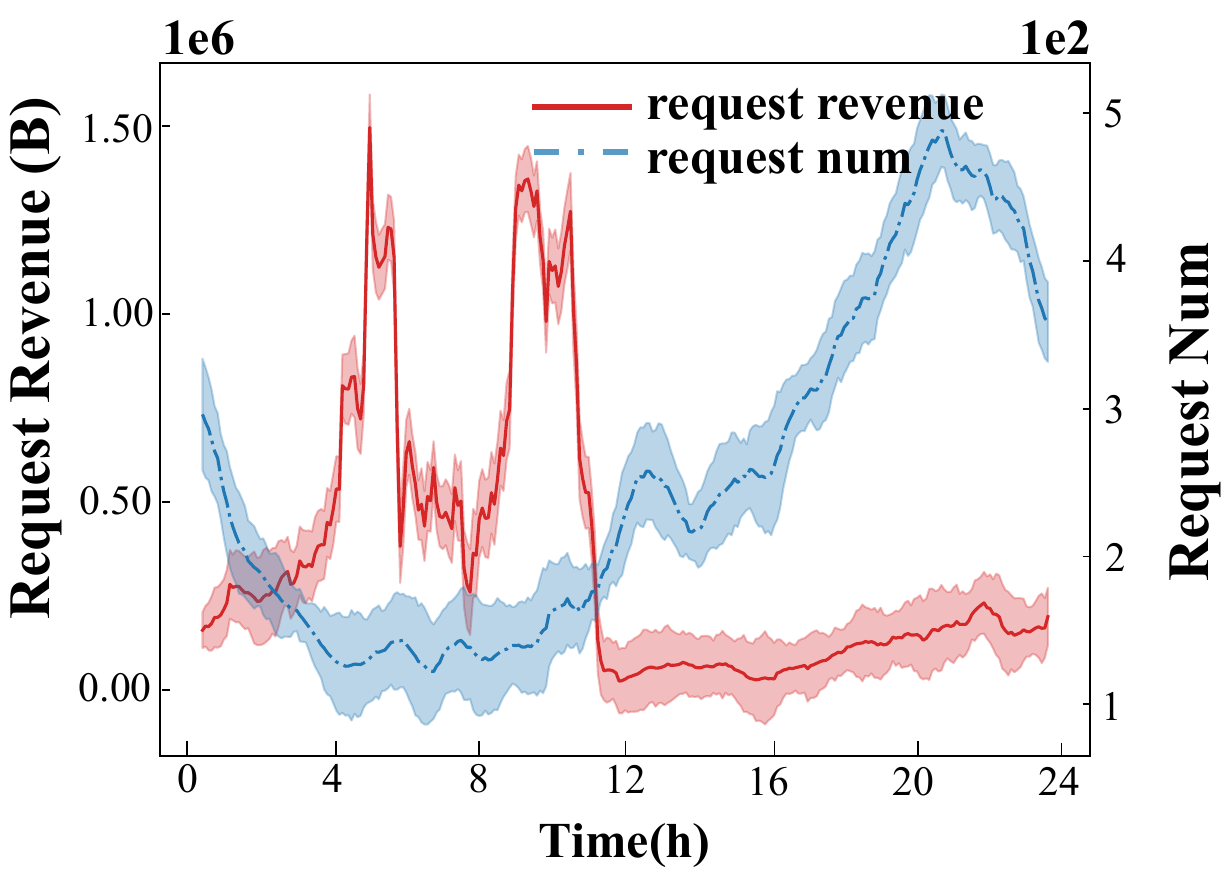} % first figure itself
        \caption{Influence of request timing.}
        \label{fig:timing}
    \end{minipage}\hfill
    \begin{minipage}{0.45\columnwidth}
        \centering
        \includegraphics[width=3.2cm]{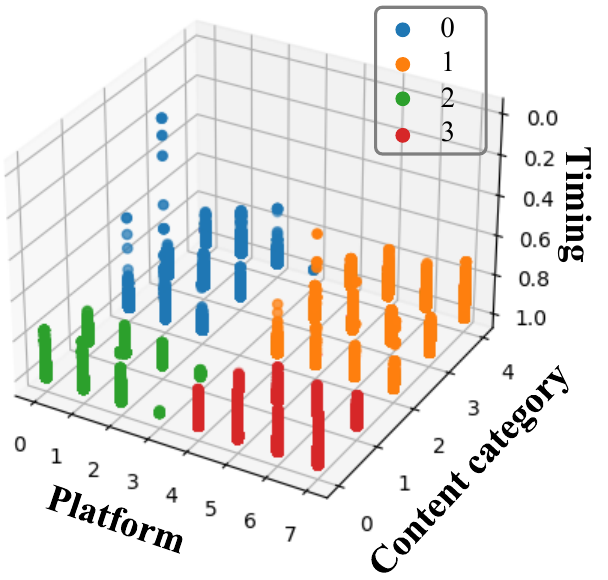} % second figure itself
        \caption{Distribution of clusters.}
        \label{fig:cluster visual}
    \end{minipage}
    % \vspace{-10pt}
\end{figure}

Lastly, we consider request timing, specifically whether the request is made during peak hours. As shown in Fig. \ref{fig:timing}, the revenue per request is typically lower when the number of requests peaks and considerably higher during times with relatively fewer requests. This trend can be attributed to users opting for lower-quality stream bitrates during peak times, and the overall decrease in  request revenue due to network fluctuations under high server load.

% 在见证了这些请求特征令人印象深刻的影响之后，我们有望在请求真正被服务前就预知请求收益（具体见第4章）。然而想要在请求预测阶段实现这一过程非常困难，因为除了请求的数量分布，我们难以同时预测请求的多种特征。为解决这一问题，我们提出利用聚类将异构特征信息融入请求的类别，即根据内容类别，平台和timing三种特征将请求聚合为多个类别，这样一来我们只需要预测每类请求的数量，而无需预测每个请求的特征，这就大大降低了模型的预测难度。

% 我们采用KMeans作为我们的聚类方法，我们认为由于各个特征内部存在明显的区分，这种基于距离的聚类方法将适用于我们的问题。图7可视化了当聚类数量K=4类时的请求点分布样例，可以看出每类请求在三个聚类指标坐标轴上都较好地被区分，同时形成了自己紧密凝聚的簇。然而我们强调K=4并不是最终选择的聚类数量，因为我们需要权衡类别数是否够多以携带足够的特征信息用于后续的请求收益预测。

% 经过Silhouette Score和Elbow Method两个指标的筛选，我们确定了对请求的最佳聚类数为K=29，

Upon observing the profound influence of these features, we are now positioned to foresee request revenue before the requests are actually served (detailed in Section \ref{sec:model_rr}). However, implementing this with the predicted requests is challenging due to the difficulty in concurrently perceiving several request features. To address this, we propose to incorporate heterogenous feature information into request categories via clustering. Specifically, we cluster requests into various categories based on content, platform, and timing, thereby allowing us to predict the volume of each category without needing to forecast individual request features, substantially simplifying the prediction task.

We adopt KMeans for clustering, given its suitability for our scenario with distinct separations across features. Fig. \ref{fig:cluster visual} illustrates the request point distribution for $k=4$ clusters. This demonstrates well-defined separation along the three axes, forming distinct and cohesive clusters. However, we stress that $k=4$ is not our ultimate cluster choice as we must ensure a sufficient number of categories to encapsulate adequate feature information for downstream request revenue prediction.

% 基于以上结论，我们采用收益（单位时间内的平均字节传输量），平台类型和是否高峰期三种特征并请求进行聚类。这样做的目的有两点：1、降低请求预测难度和模型复杂度：经过聚类后的请求可以用类收益均值来表征一类请求的收益，而不需要模型去预测每个请求的收益，这就大大降低了模型的预测难度，同时也降低了所拟合的数据分布复杂度。从而提高模型效率。2、同时建模请求的服务收益和服务成本：通过将平台类型，是否高峰期这种对QoS影响较大的特征融入请求的区分参考，我们可以建模收益以外的请求成本特征，并基于类别进行QoS量化分析来建模后续的SLA约束优化过程。

% With the insights garnered from our analyses, we perform clustering on the live streaming requests based on three distinct features. This approach serves a dual purpose: firstly, it alleviates the complexity of request prediction and model complexity. By representing the revenue of a cluster of requests with the average cluster revenue, we avoid predicting individual request revenue. Secondly, we simultaneously model the service revenue and service cost of the requests. By incorporating features that greatly influencing QoS into requests differentiation, we consider cost attributes beyond revenue and perform quantitative QoS analysis for each category, laying the groundwork for future SLA constraint optimization.

% We adopt the K-Means, deeming it appropriate given the distinct differentiation within each feature. After evaluating our clustering results using the silhouette score and Elbow Method we determine an optimal cluster number of 29 for the requests. Fig. \ref{fig:cluster visual} illustrates the resulting distribution of clustered requests, clearly demonstrating well-differentiated, tightly cohesive clusters along the three attribute axes.

\begin{figure}
    \centering
    \includegraphics[width=\columnwidth]{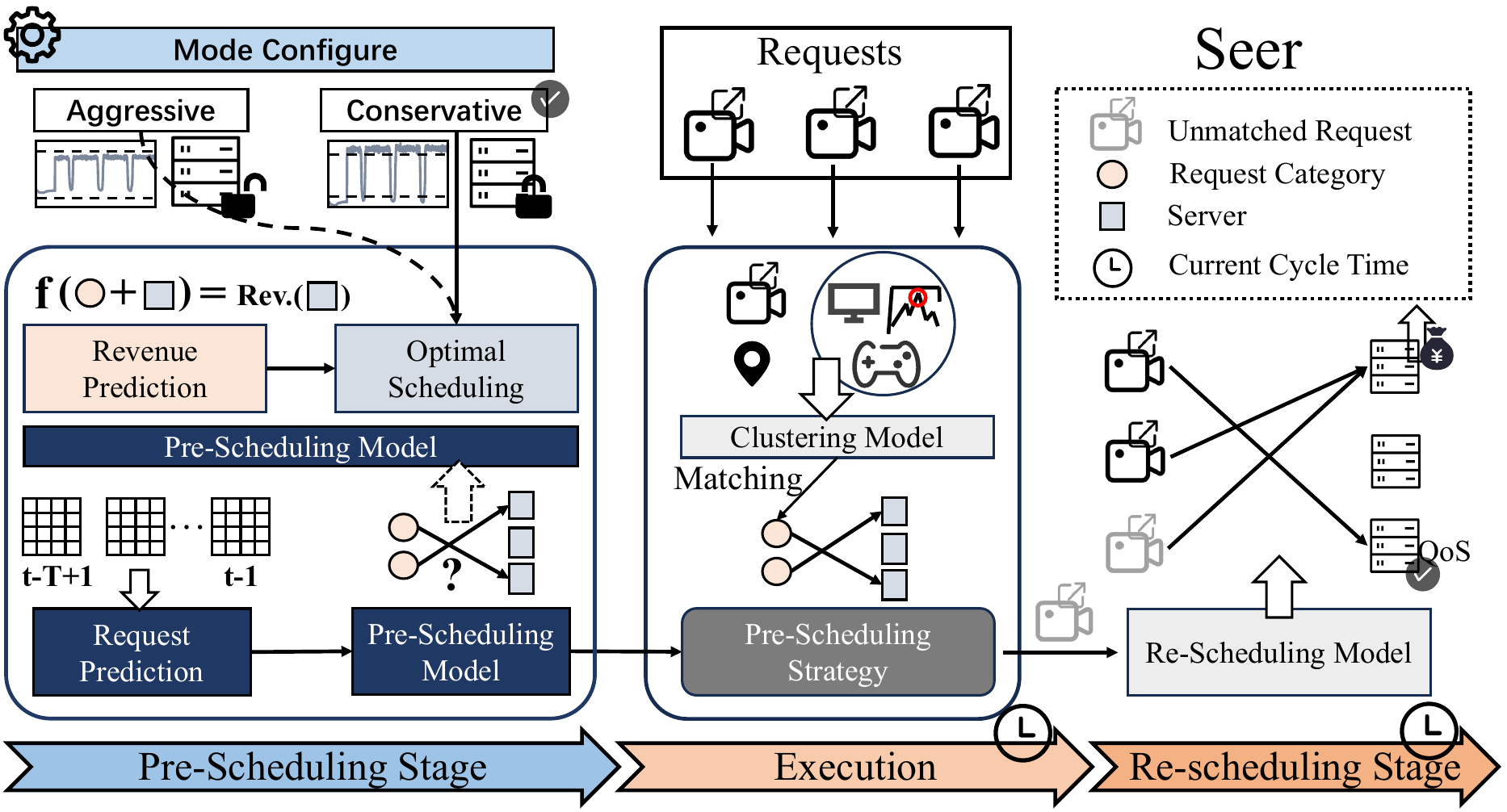}
    \caption{The system overview of Seer.}
    \label{fig:Seer}
    \vspace{-15pt}
\end{figure}

% \subsection{Takehome Messages}
% \textbf{Potential Utilization Improvement and Optimal Range:} % 通过对现实场景的分析和真实数据验证，我们发现CCP的服务器资源利用率存在较大的潜在提升空间，具体表现为超过70%的服务器的单天平均利用率低于20%，这主要是由于请求分布式时空异构性和传统CCP就近调度方案之间的冲突导致的。除此之外，我们发现服务器资源利用率与收益之间存在着复杂的非线性关系，具体表现为存在着高低利用率两个会导致收益骤减的阈值，这也启发我们为了最大化CCP的收益应该尽可能使服务器处于一个有效的收益区间。

% \textbf{Correlations between Request Scheduling and QoS:} % 

\section{SYSTEM OVERVIEW}\label{SYSTEM}
Fig. \ref{fig:Seer} provides the overview of Seer. Seer is a proactive revenue-aware live streaming scheduling system, explicitly designed to maximize CCP revenue while addressing related concerns including server withdrawal and QoS SLA. Additionally, Seer emphasizes maximizing efficiency by minimizing the proportion of decision-making time in each scheduling cycle, thereby reducing user-perceived latency caused by scheduling.

Specifically, Seer operates through a three-stage process: pre-scheduling, execution, and re-scheduling. In the pre-scheduling stage, before the actual requests arrive, Seer anticipates the category, volume, and geographic distribution of subsequent requests. This prediction enables a proactive approach to solving a revenue maximization problem under certain constraints, creating an optimal pre-scheduling strategy. 

The prediction and problem-solving stages, being the most time-consuming in Seer's execution, are expedited by completing pre-scheduling prior to actual scheduling, effectively reducing user wait times for services.
% The prediction and problem-solving processes are the most time-consuming in the Seer execution. Fortunately, by introducing prediction, we can complete the pre-scheduling prior to actual scheduling, thus reducing the time that users wait for services. 

Upon the actual scheduling cycle, Seer matches and schedules requests according to the pre-scheduling strategy, leaving only a minor proportion of requests (untreated due to prediction errors) for the re-scheduling stage, where they are managed through a simple and efficient scheduling strategy. Beyond these, Seer also employs two optional scheduling configuration modes: conservative and aggressive modes.

% Beyond these three stages, Seer also employs a default fine-tuning technique and two optional scheduling configuration schemes, namely, error feedback fine-tuning and radical/console scheduling configurations.

% Seer是一个带有CCP收益感知的直播请求调度系统，其将显式地考虑服务器逃逸与过载（即QoS SLA），将请求实时调度到最合适的边缘服务器以最大化CCP的整体收益。另外，Seer还重视调度效率的最大化，即在每轮调度中最小化决策的时间占比，减少调度导致的用户时延。

% 具体来说，Seer通过三个阶段完成对请求的实时调度，即预调度阶段，预调度方案执行阶段和重调度阶段。预调度阶段发生在每轮请求到来之前，Seer将预测下一轮请求的类别、数量和地理分布，并据此求解一个双重约束下的收益最大化问题，生成预调度方案。预测和最优化问题的求解过程在整个Seer执行过程中最为耗时，幸运的是由于预测的引入，我们可以在真实请求到来之前就完成预调度的过程，而不用占用用户等待服务的时间。在请求正式到来之时，Seer将先按照预调度方案匹配并调度用户的真实请求，而由于预测误差未被处理的少量请求将进入重调度阶段通过简单高效的调度策略进行调度。

% 除了以上三个阶段的调度，Seer还具有一种默认的微调技术和两种可选的调度配置方案，即误差反馈微调和radical/console 调度配置。

\subsection{Proactive Pre-scheduling Stage}
As the core stage of Seer, the pre-scheduling stage is composed of the following parts, as shown in the Fig. \ref{fig:pre_stage}:
\subsubsection{Request Prediction Model}

Prior to new scheduling cycle, Seer first predicts the request category distribution of all $M$ locations: $\mathcal{R}=\{R_m | m \in [1, M] \}$ from the prediction model, where $R_m=\{r_m^1,...,r_m^N\}$ denotes the number of requests of all $N$ categories in region $m$.

We introduce a proactive deep learning model that combines an AutoEncoder (AE) with a Gated Recurrent Unit (GRU) for spatial-temporal feature extraction and prediction of live streaming request matrices. The model incorporates the strength of AEs in spatial feature learning with the prowess of GRUs in capturing temporal dependencies, offering a comprehensive method to interpret spatial-temporal data.

\begin{figure}
    \centering
    % \captionsetup{belowskip=-10pt}
    \includegraphics[width=8cm]{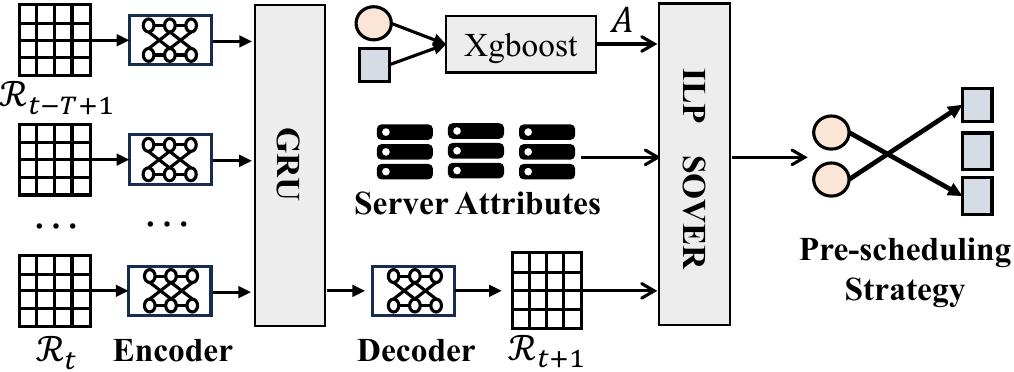}
    \caption{Process of Seer's pre-scheduling stage.}
    \label{fig:pre_stage}
    \vspace{-10pt}
\end{figure}

The architecture of the AE-GRU model (Fig. \ref{fig:pre_stage}) involves two main components. The first is the AE that learns to transform the spatial relationships across different locations and the request patterns of multiple categories into a compressed representation. The second component is the GRU, which accepts the encoded representation as input and captures the temporal dependencies between sequential matrices. The entire process of the AE-GRU can be defined as follows:
\begin{align}
\label{eq:AE}
&\mathcal{H}_t=[\mathbf{En}(\mathcal{R}_{1}),...,\mathbf{En}(\mathcal{R}_{t})] \\
&\mathcal{H}_{t+1} = \mathbf{GRU}(\mathcal{H}_t) \\
&\hat{\mathcal{R}}_{t+1} = \sigma(W_{d} \cdot \mathcal{H}_{t+1} + b_{d})
\end{align}where $\mathcal{R}_{t}$ is the input matrix at time $t$, and $\mathbf{En}(\mathcal{R}_{t}) = \sigma(W_{e} \cdot \mathcal{R}_{t} + b_{e})$ is the encoded representation. $\sigma$ represents the activation function (we use the ReLU function here), $W_{e}$ and $W_{d}$ are the weight matrices, and $b_{e}$ and $b_{d}$ are bias vectors for the AE-encoder and AE-decoder respectively. Given our use of the standard GRU, for brevity, we adopt $\mathbf{GRU}$ to represent the standard computational process inherent in the GRU.

By combining these components, AE-GRU is capable of effectively learning both the spatial and temporal features embedded in the request matrices and predicting the future request distribution. Besides, an inherent advantage of the AE-GRU is its capability of efficient feature extraction through a lightweight network structure, which is particularly valuable in real-time scheduling where every second counts.

% The AE learns a compressed spatial representation of the matrices and the GRU captures the temporal dynamics among these compressed representations, leading to accurate and efficient predictions.

% Figure 10 illustrates an instance of AE-GRU's prediction of request category distribution, alongside the actual data. AE-GRU accurately predicts live-streaming requests, thereby aiding the pre-scheduling stage reliant on the forecasted request matrix. However, like all predictive models, AE-GRU is prone to errors, particularly underperforming when predicting extreme values. These variances between predicted and real request distributions call for additional rescheduling for correction, and an update paradigm to automatically adjust model parameters in response to significant prediction errors, adapting to evolving request distributions.

% 图10展示了AE-GRU所预测的请求类别分布示例和其对应的真实情况，可以看出AE-GRU能较为精准的预测真实的直播请求，支持后续基于预测请求矩阵的预调度阶段。但我们也发现，同所有预测模型一样，AE-GRU在极端值（如最大最小值）的预测上也存在着误差，这些与真实请求分布之间的差别需要进一步的重调度来处理，同时也需要有更新机制使得模型在出现较大误差时能够自动更新其参数以适应变化的请求分布。

\subsubsection{Modeling Request Revenue} \label{sec:model_rr}
In this part, we outline our approach to foreseeing request revenue, which accounts for both request features and server attributes. As discussed in Section \ref{rev_ana}, intrinsic request features will affect the request revenue. However, server attributes, e.g., bandwidth and geographical location, also affect revenue by influencing transmission stability and speed.

To accurately model request revenue, we employ eXtreme Gradient Boosting (Xgboost), a scalable Gradient Boosted Decision Tree model. Specifically, we use request category, location, server ID, bandwidth, and location as inputs, with the corresponding request revenue as the label. Training Xgboost on ample historical data allows it to model request revenue effectively. To boost scheduling efficiency, we generate the request revenue matrix $\mathbf{A}\in \mathbb{R}^{E \times M \times N}$ in advance by iterating over all request-server combinations on the trained Xgboost:\begin{align}
\mathbf{A}_{e,m,i} = &\mathbf{Xgboost}(i, m, e, B_e, L_e)\\ 
&(e\in[1,E], m\in[1,M], i\in[1,N]) \nonumber
\end{align}where $\mathbf{Xgboost}$ represents the revenue modeling process, $i, m, e, B_e, L_e$ are corresponding input features, $E$ represents the number of servers, and $M$ and $N$ are defined as before. Xgboost is not only efficient but also generalized, aiding the of request-server combinations unseen in the historical data.
% 在这一部分我们介绍如何通过请求和服务器属性在请求被真实服务前预见请求收益，在Section \ref{rev_ana}，我们分析了请求自身的特征如何影响请求收益，并将这些信息融入到请求的类别中。然而，在现实环境中，服务器的属性也会间接地影响被服务请求的收益。例如，服务器的带宽和地理位置可能会影响到传输的稳定性和速度。如果一个服务器的带宽不足，或者地理位置离用户设备过远，可能会导致传输中断或者传输速度下降，这种情况下，请求收益也会降低。

% 为综合考虑这些因素并准确建模请求收益，我们引入了Xgboost((eXtreme Gradient Boosting), which is a scalable and efficient gradient-boosted decision tree (GBDT) machine learning model. 具体来说，我们考虑LL数据集中每条记录的请求类别，请求位置，服务器的id，带宽，位置五种属性作为输入，并将对应的请求收益作为标签，通过大量的历史数据训练Xgboost，使其能通过拟合以上特征给出对请求收益的预判。Xgboost不仅高效，其良好的泛化性可以有效帮助预测历史数据中未出现过得请求-服务器组合。

% 为提高调度效率，我们选择在训练好的Xgboost上遍历所有的请求-服务器输入组合（执行次数为E*M*N,其中E为服务器数量，M和N的定义与前文一致），提前建立请求收益矩阵$\matchbf{A}$，这样就避免了在调度过程中多次调用Xgboost。值得一提的是，随着系统的运行，历史数据不断积累，$\matchbf{A}$也需要在合适的时间做出更新，由于篇幅问题我们不重点讨论这一工程化问题。

% and generate a pre-scheduling strategy to decide which servers these requests should be distributed to. 
% non-withdrawal and QoS SLA-guaranteed % & \sum_{i=1}^N \sum_{m=1}^M x_{m,e}^i * s^i \leq \alpha \sum_{i=1}^N \sum_{m=1}^M x_{m,e}^i \ (e \in [1,E]).\\ % SLA限制 一个服务器服务的高需求请求占比小于多少
% & \sum_{i=1}^N \sum_{m=1}^M x_{m,e}^i * (L_e\oplus m) \leq \beta \sum_{i=1}^N \sum_{m=1}^M x_{m,e}^i \ (e \in [1,E]). % SLA限制. 一个服务器服务的异地请求占比小于多少

\subsubsection{Pre-scheduling Model}\label{prescheduling}With the predicted request matrix $\hat{\mathcal{R}}_{t+1}$, the core task of Seer in the pre-scheduling stage is to generate a strategy for potential requests in conjunction with the current server states, to ensure in advance that the next cycle of request scheduling is revenue-optimized.

We formally define the revenue-optimized scheduling problem: assuming that there are $E$ edge nodes and $M$ locations. The bandwidth capacity and location of server $e$ is represented as $B_e$ and $L_e$. We represent the Pre-scheduling Strategy with $PS=\{x_{m,i}^e|e\in[1,E],m\in[1,M],i\in[1,N]\}$, where $x_{m,i}^e$ denotes the number of request of category $i$ from location $m$ served by server $e$. To this end, the pre-scheduling problem at each scheduling cycle is:
\begin{align}\label{eq:object}
\max \sum_{e=1}^E \left(\sum_{m=1}^M \sum_{i=1}^N \frac{x_{m,i}^e * \mathbf{A}_{e,m,i}}{B_e}\right)
\end{align}
\begin{subequations}
\begin{align}
& \text { s.t. } x_{m,i}^e \geq 0\ (e\in[1,E], m\in[1,M], i\in[1,N]) \\  % x不为负数
& \sum_{e=1}^E x_{m,i}^e = \hat{r}_m^i\ (m\in[1,M], i\in[1,N]) \\  % 所有请求都被满足
& \alpha < \sum_{m=1}^M \sum_{i=1}^N \frac{x_{m,i}^e * \mathbf{A}_{e,m,i}}{C_e} < \beta \ (e \in [1, E]) \label{eq:mim_limit} % SLAs
\end{align}
\end{subequations}where Eq. 7 is our objective function to optimize resource utilization across the CCP, Eq. 8a imposes the constraint on the range of $x$, while Eq. 8b ensures that all requests are scheduled. Eq. 8c ensures that the utilization of each server remains within the optimal revenue range, thus avoiding server withdrawal or violating QoS SLA.

Given that excessive server workload can lead to significant QoS degradation, we count the historical QoS data in conjunction with servers' bandwidth utilization to ascertain an appropriate value for $\beta$. Initially, we calculate CDF for both QoS metrics. To ensure neither metric was excessively poor, we designate $\beta$ as the smaller utilization that corresponds to the 80th percentile in both QoS metrics:
\begin{align}
\beta = \min \{U_{Lat_{80}}, U_{Err_{80}}\}
\end{align}where $U_{Lat_{80}}$ and $U_{Err_{80}}$ correspond to the 80th percentile of the startup latency and error rates, respectively.

% The intent is to depict the proportional relationship of the utilization with the defined quantile measures for QoS.

Since $x$ in Eq. \ref{eq:object} represents non-negative integers, and both the objective function and all constraints are linear, the problem constitutes an Integer Linear Programming (ILP) problem. Although ILP problems can theoretically be solved with an analytical solution, solving ILP is NP-Hard \cite{pmlr-v139-paulus21a}. Considering the enormous scale of CCP's servers and requests, which results in a huge solution space ($E*M*N$ is 800,000+), most ILP solutions can't deliver in time.

To circumvent this issue, we propose to reduce dimensionality from the location dimension since it has less effect on request revenue. Specifically, we ignore the request location during the ILP resolution ($x_{m,i}^e\rightarrow \overline{x}_i^e$, representing the number of category $i$ requests assigned to server $e$ from all locations). We also average $\mathbf{A}$, reducing it to $\overline{\mathbf{A}}\in\mathbb{R}^{E\times N}$, representing the average revenue when server $e$ serves category $i$ requests. Consequently, Eq. 7 simplifies to:
\begin{align}
\max \sum_{e=1}^E \left(\sum_{i=1}^N \frac{\overline{x}_{i}^e * \overline{\mathbf{A}}_{e,i}}{B_e}\right)
\end{align}and the constraints of Eq. 8 can be similarly simplified. The complexity of the new ILP is reduced by $M=59$ times (from one minute to 1s or less). Moreover, we relax the integer constraint on $x$, turning the problem into a Linear Programming problem. We then use the Branch and Cut (BC) algorithm \cite{cordeau2006branch} to solve it. After obtaining the intermediate result $\overline{PS}$, we perform two processing steps to expand it:

1) Round $\overline{PS}$ to the nearest integer. 2) Expand the location dimension of $\overline{PS}$ based on the geographic distribution of $\hat{\mathcal{R}}_{t+1}$, i.e., divide $\overline{PS}\in \mathbb{R}^{E\times N}$ to $PS \in \mathbb{R}^{E\times M\times N}$ according to the proportion of request in different locations, ensuring:
\begin{align}
\frac{x_{m,i}^e}{\overline{x}_{i}^e} = \frac{\sum_{i=1}^N \hat{r}_m^i}{\mathbf{SUM}(\hat{\mathcal{R}}_{t+1})}
\end{align}where $\mathbf{SUM}(\hat{\mathcal{R}}_{t+1})$ represents the number of requests. Despite the risk of reducing expected revenue due to the simplification, we deem this step essential, as Seer needs to ensure the completion of pre-scheduling before actual requests arrive.

\subsection{Pre-strategy Execution Stage}\label{pre-s execution}
\begin{figure}
    \centering
    % \captionsetup{belowskip=-10pt}
    \includegraphics[width=7cm]{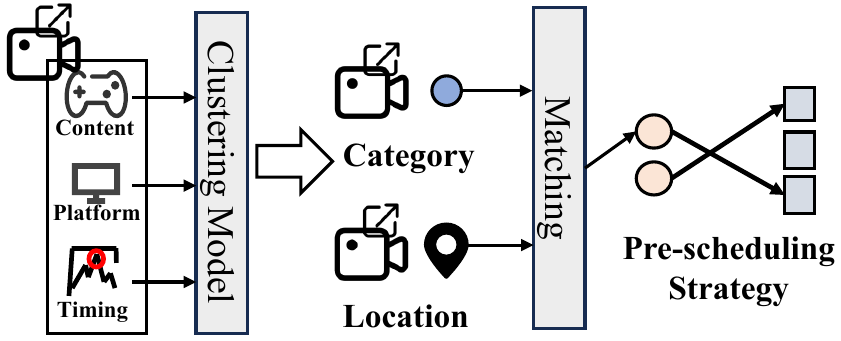}
    \caption{The process of execution stage.}
    \label{fig:execution}
    \vspace{-12pt}
\end{figure}

Once the $PS$ is constructed, Seer proceeds to reconcile the real-time incoming requests with the preconceived strategy and schedules them accordingly (as illustrated in Fig. \ref{fig:execution}).

With the arrival of actual requests, Seer classify them into predefined categories by the clustering model. Then, for each respective location and category, the requests are dispatched to servers in accordance with the pre-scheduling strategy.

The request matching can be formalized as $f(\mathcal{R}_{t+1}, PS) \rightarrow S$, where $f$ is the matching function, $\mathcal{R}_{t+1}$ is the actual request matrix, and $S=\{s_{m,i}^e|e\in[1,E],m\in[1,M],i\in[1,N]\}$ is the adapted scheduling strategy. Specifically, $f$ allocates requests of category $i$ from location $m$ to server $e$, according to the magnitude and order of $x_{m,i}^e$ in $PS$, ensuring that servers with greater serving capacity $x_{m,i}^e$ are satisfied first.

Additionally, as actual requests deviate from the prediction, leading to a discrepancy between the number of requests and the number allocated in $PS$. To counter this, Seer adopts a \emph{request priority} principle, whereby if requests exceed the allocated quantity, the surplus is logged for subsequent handling. Conversely, if requests are fewer, servers that were not assigned the requisite number of requests are ignored, ensuring that $\sum_{e=1}^E s_{m,i}^e \leq r_m^i$. The unmatched requests will be further handled in the re-scheduling stage.

% Seer employs a request priority principle. Namely, if the actual requests exceed the allocated amount, the surplus is separately recorded and handled subsequently. Conversely, if fewer than allocated, servers not assigned the expected number of requests are ignored, ensuring $\sum_{e=1}^E s_{m,i}^e \leq r_m^i$.

% 真实请求的匹配过程可以被形式化为$f(\mathcal{R}_{t+1}, PS) \rightarrow S$，其中$f$代表匹配过程，$\mathcal{R}_{t+1}$代表真实的请求矩阵，$S=\{s_{m,i}^e|e\in[1,E],m\in[1,M],i\in[1,N]\}$代表适配真实情况后的调度策略。具体来说，$f$将$\mathcal{R}_{t+1}$中第m个地区的第i类请求，按照$PS$中$x_{m,i}^e$的大小h和顺序分配给服务器e，即保证对特定请求服务能力较大的服务器先被满足。

% 另外，In the event that actual requests deviate from the prediction,可能会导致真实请求数量不等于PS中所分配的请求数，为解决这一问题，Seer采取请求优先原则，即当真实请求多于分配数量时将多出的请求单独记录并在后续处理，当少于分配数量时，忽视未被分配应有请求数量的服务器。保证$\sum_{e=1}^E s_{m,i}^e \leq r_m^i$

% We define the categories as $C = {c_1, c_2, \ldots, c_n}$, the servers as $\mathcal{E} = {e_1, e_2, \ldots, e_E}$ and the incoming request vector of category $i$ as $r_i = r_1^i,r_2^i,...,r_m^i$. The allocation of requests is governed by a function $f : C \times \mathcal{E} \rightarrow [0, 1]$ which is derived from the pre-scheduling strategy $S$, such that the requests allocated to server $e$ under category $i$ are $R_{i, e} = \phi(i, e) \cdot r_m^i$.

\subsection{Re-scheduling Stage}\label{rescheduling}
In this stage, Seer assigns the remaining unserved requests according to CCP's original heuristic scheduling method, which schedules requests to the servers that is closest and has the most remaining bandwidth. The remaining bandwidth, $B_e^{remain}$, is computed using the updated scheduling strategy $S$, and revenue matrix $A$, such that $B_e^{remain} = B_e - \sum_{m=1}^M \sum_{i=1}^N s_{m,i}^e*A_{e,m,i}$. Upon completing this heuristic scheduling, Seer successfully schedule all actual requests.

Essentially, Seer operates cyclically, beginning with pre-scheduling, progressing to execution and re-scheduling, and then starts the next cycle based on the revised server state and requests. Given that Seer's execution and re-scheduling stages involve minimal computations, it allocates considerable time for the next-cycle's pre-scheduling. 

Notably, throughout long-term operation, Seer will update its components and parameters, including the clustering model, the request and revenue prediction models, and the QoS utilization $\beta$.

\begin{figure*}[ht]
\centering
\subfigure[The scheduling revenue comparison]{\label{fig:rev_compare}
\includegraphics[width=5.7cm]{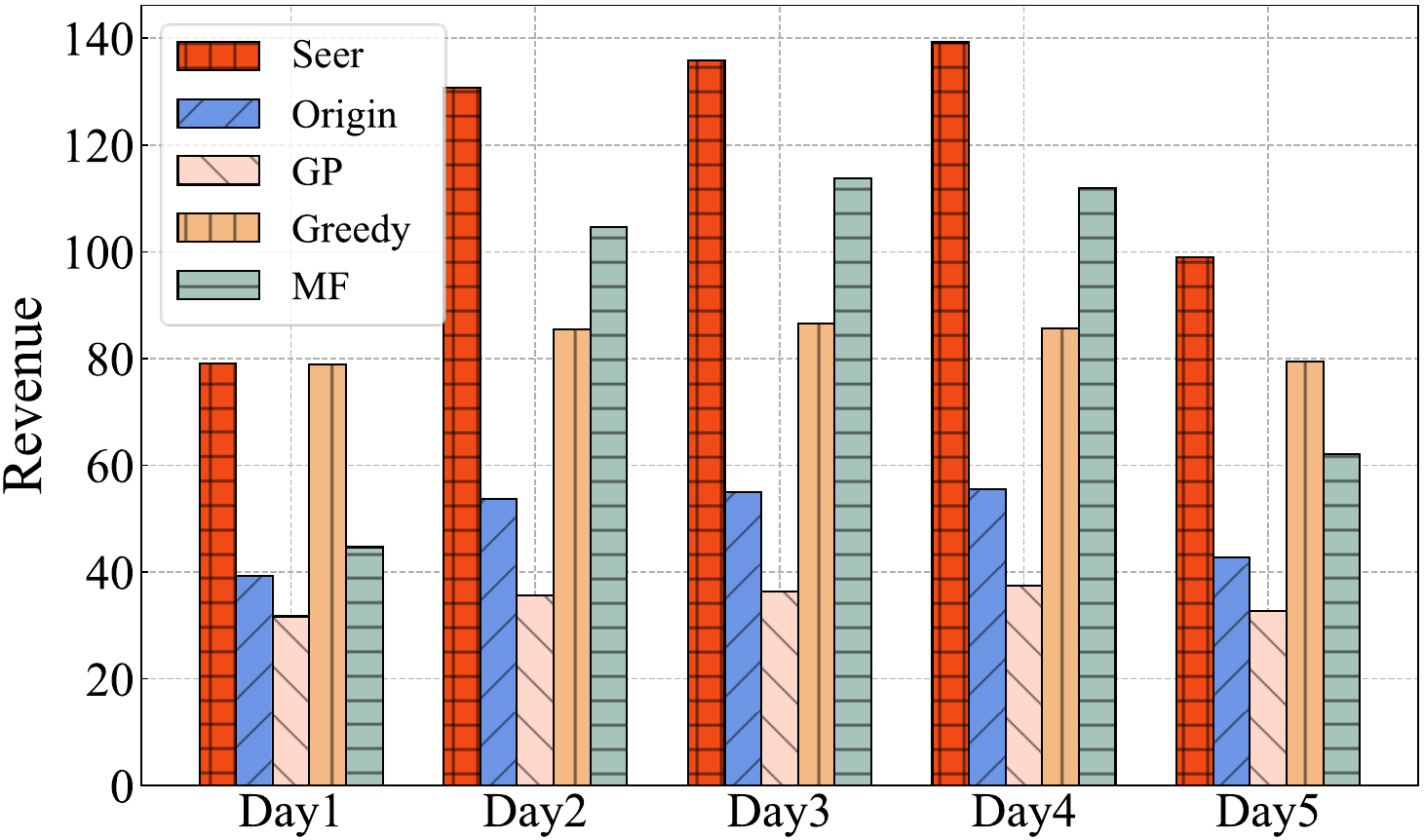}
}
% \hspace{-3mm}
\subfigure[The average utilization comparison]{\label{fig:util_compare}
\includegraphics[width=5.7cm]{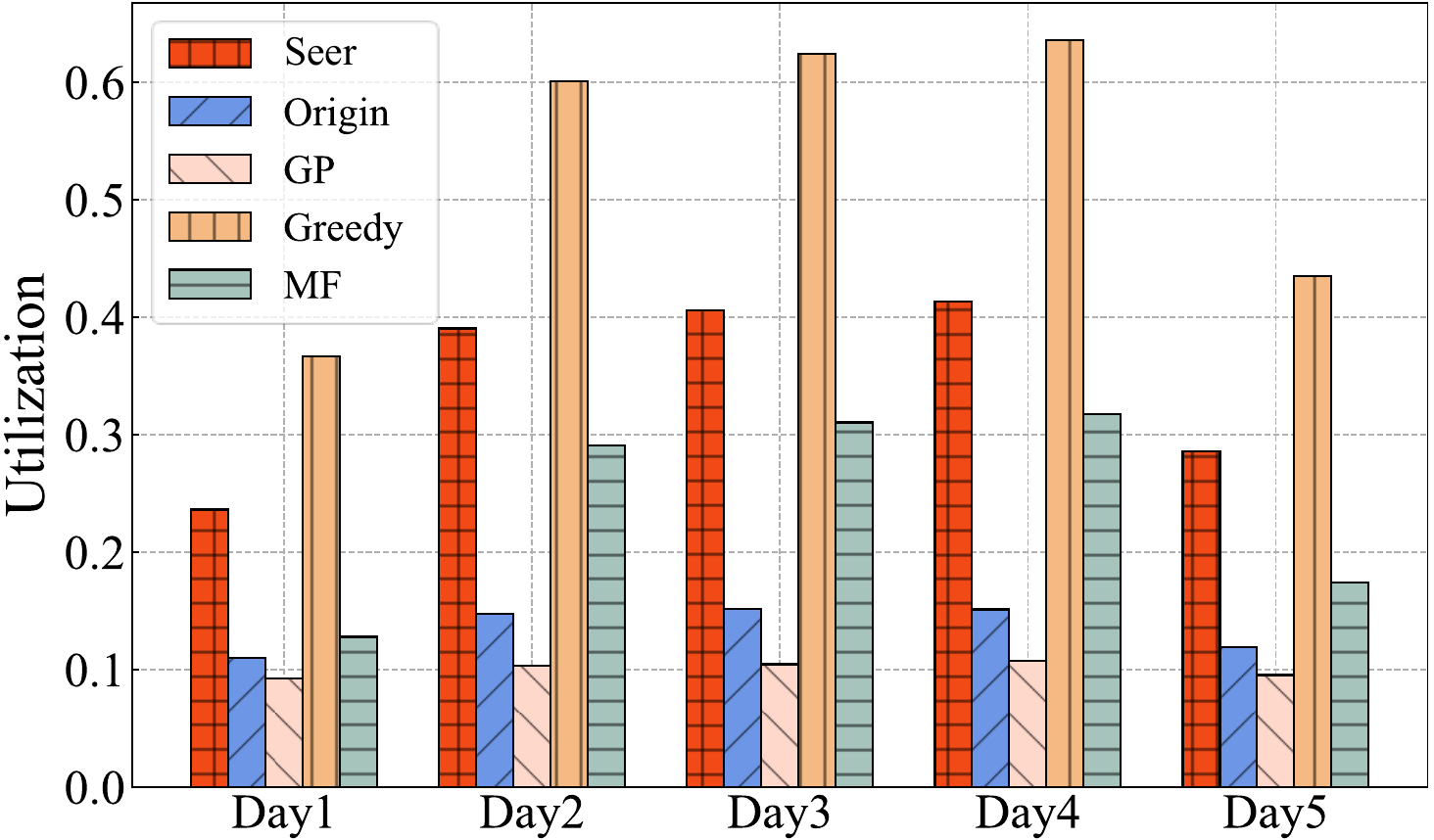}
}
\subfigure[Server withdrawal ($\alpha$) and SLA violations ($\beta$)]{\label{fig:sla_compare}
\includegraphics[width=5.7cm]{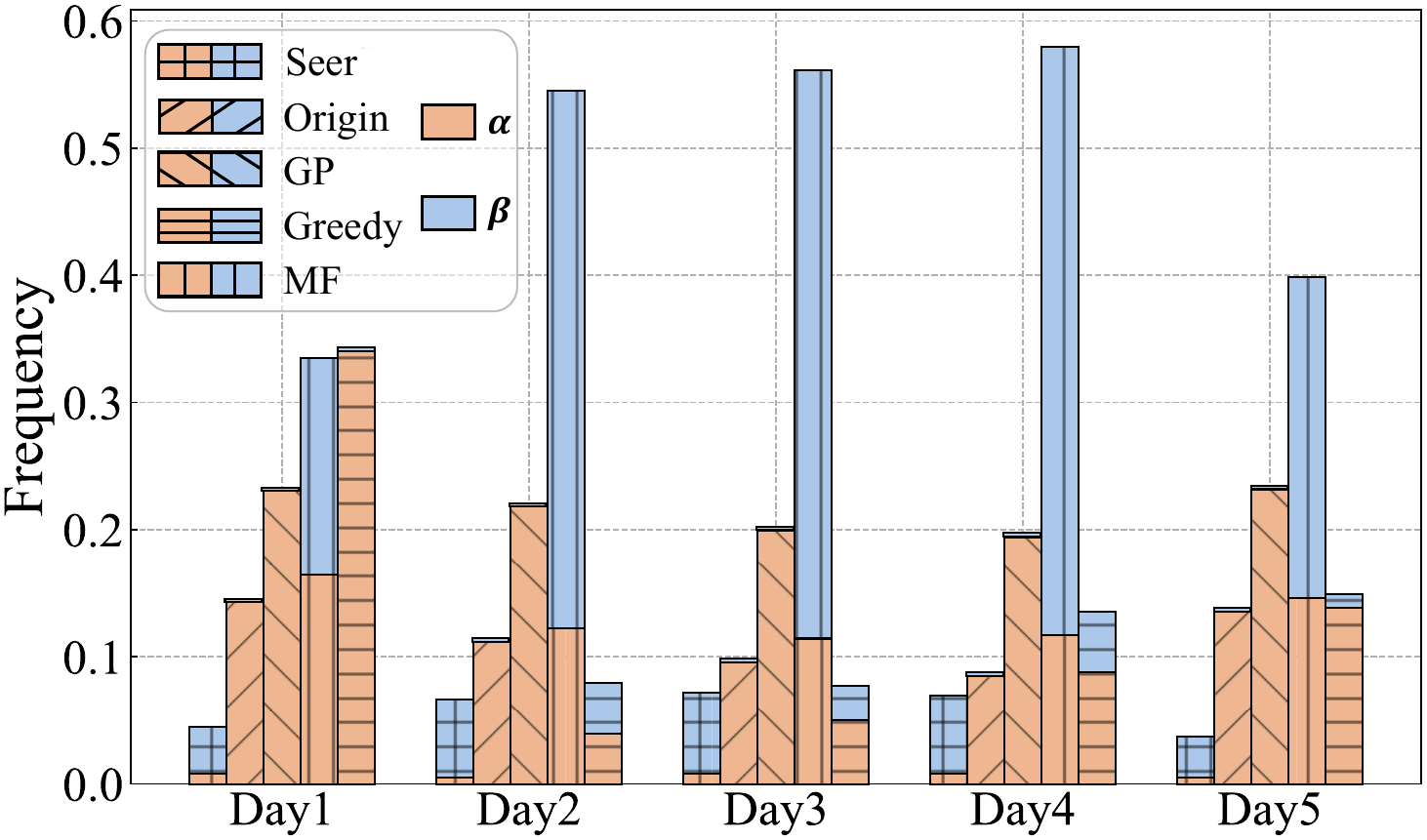}
}
\caption{Scheduling performance comparison\protect\footnotemark[1].}
\label{fig:Revenue-SLAs-Efficiency}
\vspace{-10pt}
\end{figure*}
\footnotetext[1]{As the time comes to day2 (weekend), the scale of requests increases significantly.}

\subsection{Flexible Scheduling Model}
We introduce two modes for Seer: Conservative and Aggressive modes. The previous settings describe Seer's default conservative mode, which prioritizes RPs' engagement and prevents server withdrawal. However, under certain circumstances, the more lucrative Aggressive mode is preferable for higher CCP revenue and lower server operational costs, despite potential server withdrawal. This mode modifies Seer's pre-scheduling and re-scheduling stages. During pre-scheduling, Seer discards the minimum utilization constraint in Eq. \ref{eq:mim_limit} and ignores servers with utilization $< \alpha$ in Eq. \ref{eq:object}. During the re-scheduling, Seer checks the potential server utilization, discards all servers with utilization $<\alpha$, and reallocates their requests to the remaining servers with the lowest utilization.
% We introduce two scheduling modes for Seer: the Conservative and the Aggressive modes. The previous settings describe Seer's conservative mode, which prioritizes RPs engagement and seeks to prevent server withdrawal, typically aligning with our collaborating CCPs' requirements. However, under certain circumstances, a more lucrative mode (Aggressive mode) is preferable for higher CCP revenue and lower potential server operational costs, even at the cost of possible server withdrawal.

% Specifically, the Aggressive mode changes Seer's pre-scheduling and re-scheduling stages. During pre-scheduling, Seer discards the minimum utilization constraint in Eq. \ref{eq:mim_limit} and ignores servers with utilization $< \alpha$ in Eq. \ref{eq:object}). During the re-scheduling, Seer iteratively checks the potential utilization of all servers, discards all servers with utilization $<\alpha$, and reallocates their requests to the remaining servers with the lowest utilization.

\section{PERFORMANCE EVALUATION}\label{EVALUATION}

\subsection{Experiment Setup} \label{exp_setting}
\subsubsection{Implementation Details}
% 我们对Seer的实现和评估建立在第三章所提到的三组真实数据集上，具体来说，实验涉及到2023年4月11日到21日10天3亿+用户直播请求,分布在59个不同的地区，在这期间CCP用于服务的边缘服务器数量为376个。我们使用前5天的数据完成第III章的系统测量和验证，以及第IV章对Seer的初步构建。包括验证潜在的利用率提升，统计流失利用率阈值$\alpha=0.05$，构建QoS指标到QoS利用率阈值$\beta$的映射函数函数，构建请求聚类模型RCM（类别数K=4）,验证请求分布的时空相关性并构建请求预测模型RPM,建立收益预测模型（RPM）等。

The evaluation of Seer is based on simulated scheduling of real requests from the three real datasets discussed in Section \ref{npd}. The whole process ensures that the scheduling mechanism is anchored on actual CCPs scheduling situation. Specifically, our experiments involve processing over 500
million live-streaming requests spanning 10 days, from February 5th to 14th, 2023, and the data is selected for covering a complete natural week to demonstrate the model's performance under different magnitude fluctuations.

% These requests were distributed across 59 different geographical regions and were served by 376 edge servers owned by CCP during the study period.

The data from the first five days are utilized to carry out the CCP measurements and verify the setup of Seer, and the last five days are used to test the scheduling performance. Based on the analysis in Section III-A, we set the request cluster number $k=29$, $\alpha=0.05$, and $\beta$ fluctuates in 0.7 to 0.9 with the operation of the CCP. Besides, we calculate the server revenue with a utilization of $U$ by the following function:
\begin{align}  % \alpha和\beta，聚类数量K，
    \text{rev}(U) = \begin{cases}
0, & U < \alpha, \\
0.2U, & U > \beta, \\
U, & \text{otherwise.}
\end{cases}
\end{align}The setting of $0$ and $0.2U$ is follow the analysis of anomaly penalties in Section \ref{3b}. To simplify the model, we disregard the potential price variances across different RPs and LSPs. The decision interval of Seer is set as 1 min, which is intended to be consistent with real-world CCP settings.

% Including validation of potential utilization improvements, determination of the utilization threshold $\alpha$ at 0.05, development of a mapping function from QoS metrics to the QoS utilization threshold $\beta$, establishment of the Request Clustering Model (RCM) with 29 categories ($K=29$), and verification of the spatial-temporal correlation of request distribution for the construction of the Request Prediction Model (RPM).

%Since the number of CCP servers is much larger than the number of locations and source servers for LSPs, we assume that each CCP server can serve requests for arbitrary content. % 由于CCP服务器的数量远大于所服务的地区数和LSP的源服务器数量，我们假设每台CCP服务器都可以服务对任意内容的请求。

\subsubsection{Baselines}
\begin{itemize}
    \item Heuristic method (Origin): This method schedules requests to the nearest servers with maximum remaining bandwidth. Its the origin scheduling method of our collaborated CCP, and we do not perform additional replication but only record the relevant metrics under the real CCP.
    \item Geographically-Proximate (GP): This approach schedules requests to the nearest available  edge servers. It is the most prevalent method in industry. % 启发式算法为每个请求选择距离最近且剩余带宽资源最多的服务器来进行服务，这一方法在工业界也较为流行。
    \item Revenue-aware Greedy (Greedy) \cite{Haouari2019}: This method is based on the request revenue matrix $\mathbf{A}$. It allocates requests to the server that yields the highest revenue until the server's bandwidth capacity is reached. It can be regarded as the upper-bound in the context of utilization.%贪心调度基于第三章所构建的收益预测模型并将请求分配给产生收益最高的服务器直到达到服务器的带宽上限。
    \item Maximum-flow (MF) \cite{9488868, 8241883}:. This kind of algorithms convert the scheduling problem into a flow control problem (the server is treated as the graph node, the constrained bandwidth is treated as the link capacity, and the scheduled request revenue is treated as the flow).
\end{itemize}

All models undergo evaluation on a unified scheduling decision server, handling requests of the same scale. This setup guarantees a fair and equitable comparison between different models while ensuring that the time consumption of each algorithm is consistently measured and comparable.

\subsection{Scheduling Performance}  % 调度效果说明
\subsubsection{CCP Revenue}  % 对比收益
We conduct continuous scheduling over the last five days and compare the performances of the Seer and other baselines. As can be seen in Fig. \ref{fig:rev_compare}, Seer consistently outperforms all baselines across the five testing days, showcasing superior revenue due to its optimized scheduling policy that expertly manages server resources within the optimal range for maximizing revenue.

Specifically, compared to the original heuristic method, Seer manifests an enhancement of 147\% in the average revenue. Against the GP method, the improvement of Seer is even more substantial, almost $4\times$ the GP revenue. Though falling marginally short of Seer (30\% revenue reduction), the MF method still outperforms the other approaches.

Regarding average resource utilization (Fig. \ref{fig:util_compare}), the Seer achieves the highest reasonable utilization (second only to Greedy's upper-bound). The Greedy method exhibits a higher utilization owing to its inherent scheduling mechanism that favors servers with the highest potential revenue. However, the non-linear relationship between revenue and utilization often leads to the Greedy method surpassing the QoS utilization $\beta$, thereby diminishing its revenue.

Examining schedule principles, the Greedy and GP methods focus on short-term or local optimization, overlooking overall resource allocation and revenue-related constraints. In contrast, the Seer and MF methods consider the broader network state, yielding more balanced utilization and higher revenue. 

% Notably, the Seer further integrates future information and revenue-aware considerations into present decisions, effectively preempting  potential servers withdrawal and overload situations, which ultimately leads to its superior performance.

\subsubsection{Server Withdrawal and QoS SLA}
Fig. \ref{fig:sla_compare} shows the server withdrawal and QoS SLA violations frequency of different methods' scheduling. In terms of the combination of those both metrics, Seer demonstrates clear advantages by consistently optimizing scheduling to maintain server utilization within the optimal revenue range (i.e., $\alpha<U<\beta$).

Analyzing the frequency of server withdrawal ($\alpha$ in Fig. \ref{fig:sla_compare}), the Origin and GP methods demonstrate higher withdrawal, up to 23\% for GP and 14\% for Origin, due to their conservative nature which results in extremely low utilization. 

The Greedy method shows a reversal trend, demonstrating the highest QoS SLA violations (up to 46\%) since it locally optimizes the utilization. Meanwhile, the MF's performance fluctuates with varying request scales. Notably, at smaller scales, MF is prone to server withdrawal, and inversely, it's susceptible to QoS SLA violations at larger scales.

Despite the superiority of Seer, it is not entirely immune to server withdrawal and QoS SLA violations due to the existence of prediction errors and the approximation in the pre-scheduling stage. However, these occurrences are minimal and reflect the balance achieved by the Seer between maximizing CCP revenue and ensuring system efficiency.

\subsubsection{Scheduling Efficiency}  % 调度效率
\begin{figure}
    \centering
    \includegraphics[width=6cm]{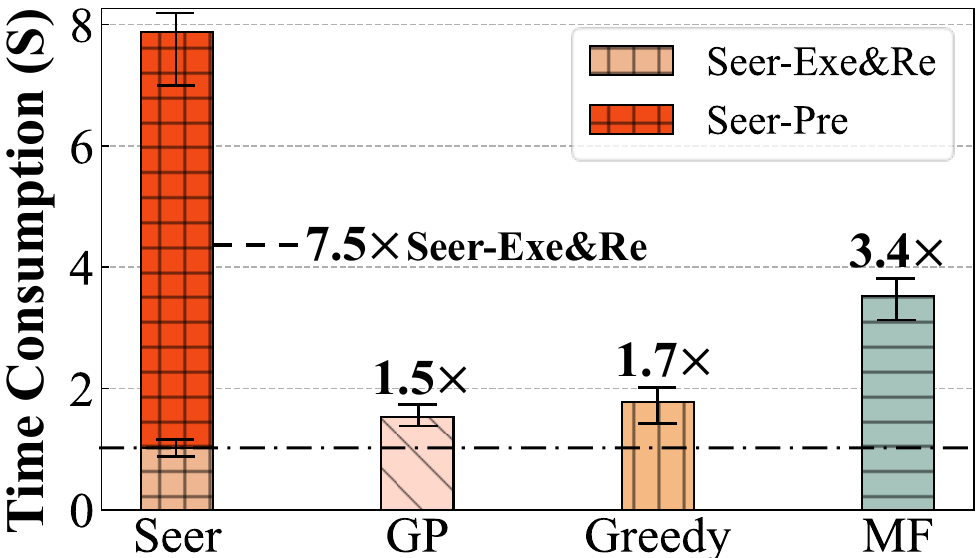}
    \caption{Average scheduling time consumption.}
    \label{fig:time_compare}
    \vspace{-12pt}
\end{figure}

We compare the average time consumption for each scheduling cycle of different methods. As illustrated in Fig. \ref{fig:time_compare}, Seer exhibits impressive efficiency in scheduling, attributable to its unique prediction and \textbf{PER} paradigm. Seer strategically shifts the time-consuming optimization problem-solving step to before the arrival of real-time requests, leaving the time-efficient execution and re-scheduling stages for the actual scheduling cycle. Seer reduces the time consumption for scheduling by $1.5\times$, $1.7\times$, and $3.4\times$ compared to GP, Greedy, and MF, respectively.

It's notable that existing state-of-the-art scheduling methods (e.g., \cite{zhang2019livesmart, 9047133}), don't fully leverage the predictive advantage. They solve the optimization problem based on real-time requests (similar to Seer's pre-scheduling stage). This choice notably increases user-perceived latency. For example, if Seer conducted pre-scheduling during actual request arrivals, the total scheduling time would inflate by $7.5 \times$, despite we have already simplified the optimization process.
% 值得一提的是，现有的stae-of-the-art的调度方法（例如xx），由于没有充分利用预测带来的'先知'优势，而是选择基于实时到来请求信息去求解最优化问题（类似Seer在预调度阶段所做的事），这将大幅提升用户因调度所感知到的时延，举例来说，如果Seer的预调度过程发生在真实请求到来时，其整体的调度时间将会增加7.5倍，还是在已经充分简化了最优化求解过程的前提下。

% In summary, the superior efficiency of the Seer model in scheduling time, combined with its ability to conduct pre-scheduling before the formal request arrives, showcases its practical advantages and applicability in real-world situations. The method presents a robust balance between optimization and scheduling efficiency, thus promising improved service delivery in the CCP scenario.

\subsection{Discussion of Generalizability}  % 真实场景的泛化性讨论
\subsubsection{Differences in Scheduling Modes}  % 对比两种调度模式
\begin{figure}
\centering
\subfigure[General metrics]{\label{fig:mode1}
\includegraphics[width=4.0cm]{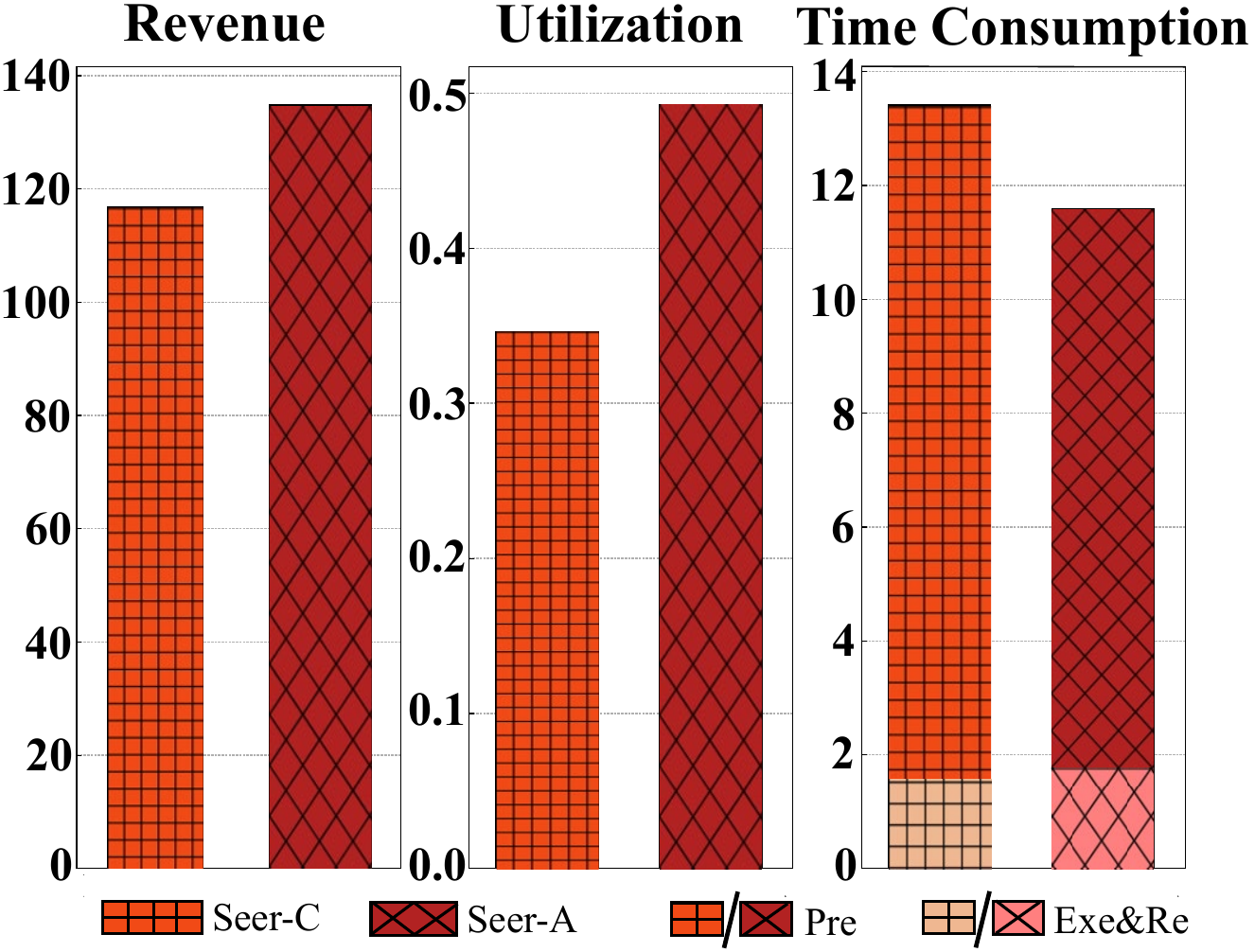}
}
\hspace{-3mm}
\subfigure[Servers utilization CDF]{\label{fig:mode2}
\includegraphics[width=4.4cm]{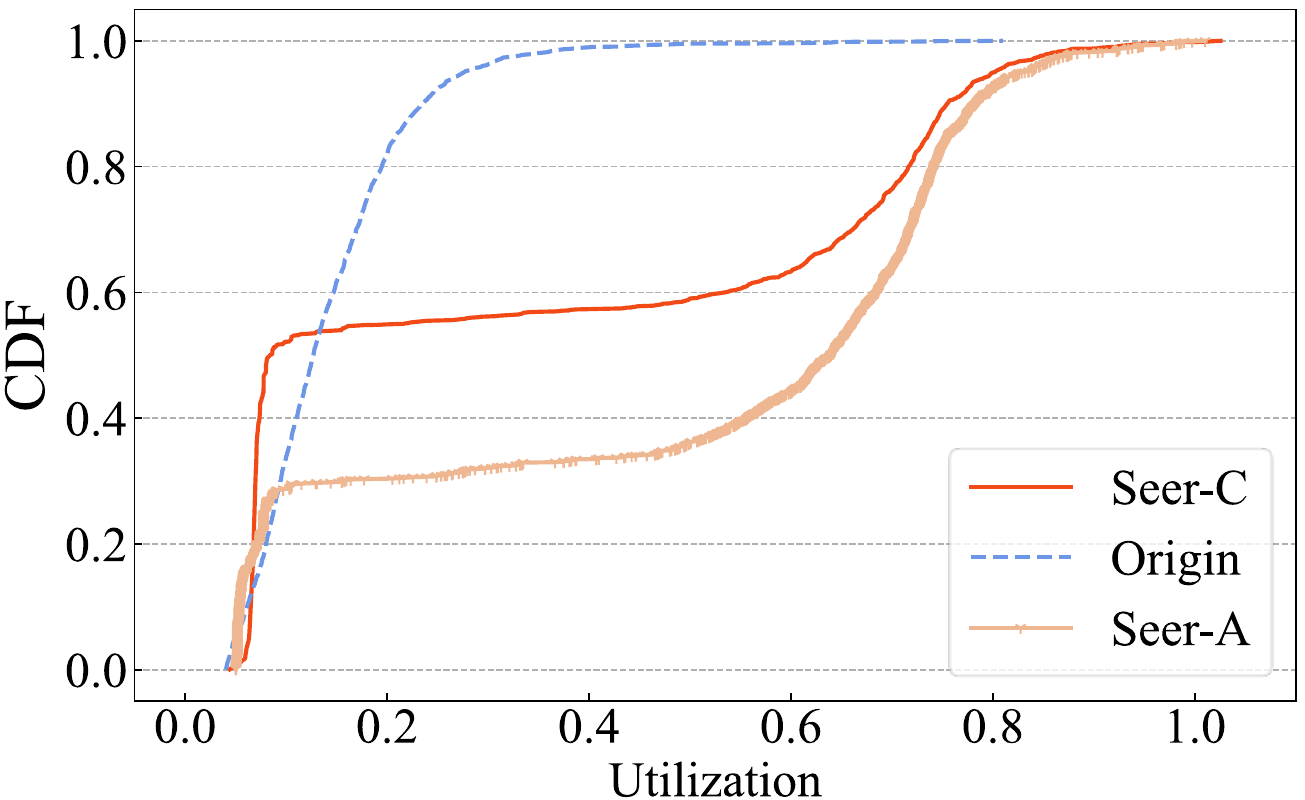}
}
\caption{Comparison of different scheduling modes.}
\label{fig:mode}
\end{figure}

To compare the two scheduling modes of Seer, we implement both Seer-C (Conservative mode) and Seer-A (Aggressive mode) on the five-day dataset. As illustrated in Fig. \ref{fig:mode1}, Seer-A outperforms Seer-C in generating higher revenue and better resource utilization, and due to the reduction of constraints in the optimization problem, it consumes less time in the pre-scheduling stage but more during the actual scheduling cycle. Fig. \ref{fig:mode2} shows how, in contrast to the origin scheduling strategy, Seer reshapes the resource utilization across CCP servers, optimizing resource use more effectively, which is more evident in Seer-A.

Thus, if CCP owner can tolerate sacrificing some loss in RPs engagement (with Seer-A discarding, on average, 33\% of servers) in exchange for enhanced gains, Seer-A often presents a more appealing revenue proposition. However, it does demand trade-off long-term operational balance within the LSPs, CCP, and RPs ecosystem, ensuring an adequate supply of cooperative RPs.

%为了对别AggCast的两种调度模式，我们在同样的五天测试数据上运行了Seer-C(Conservative)和Seer-A(Aggressive mode),图12a展示了两种模式的Seer的调度平均收益，利用率和时间消耗 。可以看出，Seer-A相比Seer-C获得了更高的收益和更资源利用率，并且由于其减少了最优化问题的限制项，其在预调度阶段所消耗的时间更少，而在真实调度轮次中消耗了更长的时间。从图12b可以看，相比于原始调度方案，Seer重构了CCP服务器的资源利用率分布，使更多服务器的资源被充分利用，Seer-A的结果中这一结论会更加明显。

% 因此我们说明，如果CCP的决策者们可以接受牺牲一些RPs的参与（Seer-A平均会丢弃33%的服务器）来换取更高的利益时，选择Seer-A往往能提供更具吸引力的收益，但这也需要CCP权衡长期运营下LSPs,CCP,RPs三者所组织成的整体产业生态，以防出现找不到可以合作的RPs的情况。

\subsubsection{Effect of Real-world Parameters}  % 对比一些真实世界中可变的参数
\begin{figure}
    \centering
    \includegraphics[width=4.5cm]{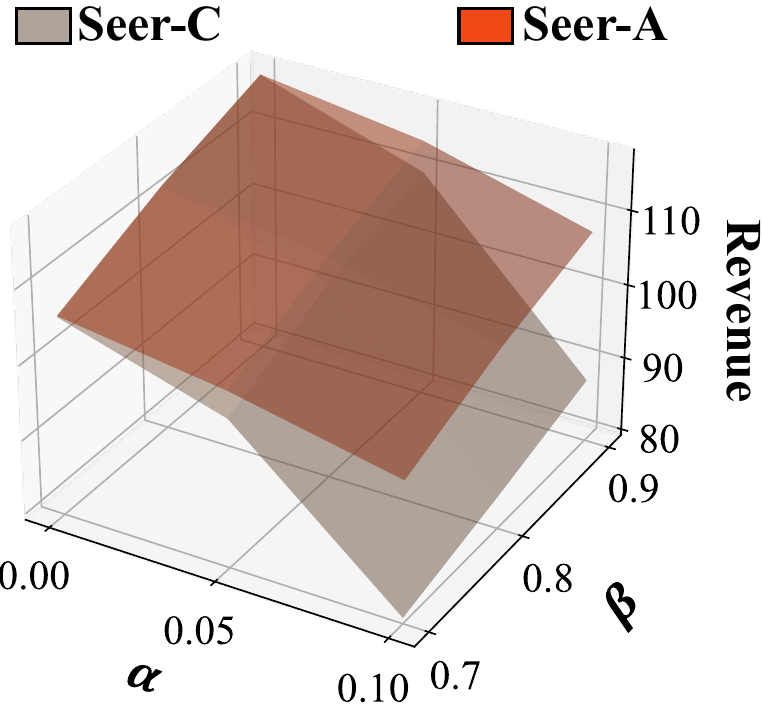}
    \caption{Variation of revenue with utilization threshold.}
    \label{fig:3d_compare}
    \vspace{-10pt}
\end{figure}

Fig. \ref{fig:3d_compare} depicts the influence of varying $\alpha$ and $\beta$ thresholds on Seer's revenue. As $\alpha$ increases, signifying RPs' heightened utilization demand, Seer's scheduling scope shrinks, subsequently diminishing CCP's revenue. Conversely, a rise in $\beta$ gradually improves Seer's revenue as it indicates enhanced network condition and hardware configurations, enabling the servicing of more requests while maintaining QoS. It's worth noting that when $\alpha = 0$, both modes of Seer yield identical revenue. Beyond this instance, Seer-A consistently outperforms Seer-C, given no servers are discarded when $\alpha =0$, causing Seer-A to revert to a strategy identical to Seer-C.

The findings above suggest that CCPs could be proactive in negotiating with RPs or offering incentives to enhance their engagement (lowering $\alpha$), and optimize the overall network environment and hardware configurations to increase $\beta$. Consequently, CCPs would have a broader scheduling scope to generate greater revenue with limited resources. We argue that such a discussion about this trade-off is of immense value as it contributes to the prosperity of the live streaming ecosystems.

% 图13展示了Seer在不同利用率阈值$\aplha$和$\beta$设定下的收益表现，可以看出，随着$\alpha$增加，RPs对所提供服务器的最低利用率要求增加，CCP的调度空间被压缩，Seer的收益也随之减少。另一方面，$\beta$的增加则逐步提高了Seer的调度收益，这是因为$\beta$增加意味着网络状况和硬件配置提升，可以在保证QoS的前提下服务更多请求。另外可以看到当$\alpha = 0$时两种模式下的Seer调度收益相同，出这种情况外Seer-A都展现出了比Seer-C更优异的收益表现，这是由于当$\alpha =0$时不存在需要丢弃的服务器，此时Seer-A将退化成与Seer-C完全相同的调度策略。

% 通过对现实世界中影响收益的参数分析，其结果鼓励CCP积极与RPs进行协商或给与激励以提高其参与热情（从而降低退出阈值$\alpha$），同时可以通过优化平台的整体网络环境和服务器硬件配置来提高QoS阈值$\beta$，这样一来CCP将拥有更大的调度空间在有限的资源上创造更多的收益。我们认为对这一权衡的探讨非常有价值，因为这利于整个直播产业生态的繁荣。

\section{Conclusion}
This paper introduces Seer, a proactive revenue-aware scheduling system for live streaming services in CCPs. Inspired by our meticulous measurements of real-world CCP environments, by integrating a novel PER paradigm and flexible scheduling modes, Seer exhibits efficient revenue-optimized scheduling in large-scale CCPs. Over five days of testing, Seer demonstrated superior performance, boosting CCP origin revenue by 147\% and accelerating scheduling $3.4\times$ faster than its counterparts.

% This paper introduces Seer, a proactive revenue-aware scheduling system for live streaming services in CCPs. The design of Seer is motivated by our meticulous measurements of a large-scale CCP production environment, based on which we achieves resource utilization-revenue modeling and overcoming three principal obstacles that hamper the integration of accurate prediction and optimized scheduling. With a novel PER paradigm and flexible scheduling modes, Seer exhibits efficient revenue-optimized scheduling in large-scale CCPs. Through five days of test scheduling, Seer substantially outperforms competing scheduling methods in terms of revenue, utilization, server withdrawal, and SLA violation, exhibiting a 150% improvement in CCP revenue compared to conventional methods, and delivering scheduling acceleration that is 3.4 times faster than its competitors.

%本文提出了Seer, a proactive revenue-aware scheduling system for live streaming services in CCP. Seer的设计驱动自我们对大规模CCP生产环境细致入微（或者说抽丝剥茧）的测量，基于此我们成功建模了资源利用率与CCP收益的关系，并排除了阻碍精确预测与优化调度有机结合的三个主要障碍。Seer拥有一个新颖的PER调度机制和灵活的调度模式，赋予了它在大规模CCPs下实现高效利益优化调度的能力，在5天的测试调度中，Seer在收益、利用率、服务器退出和SLA违反上都表现出了远超其他调度方法的能力，相比于原始的调度方法，RevCast提升了150%的CCP收益并实现了比优竞争对手3.4倍的调度加速。
% that combines request prediction and scheduling optimization through a novel Pre-scheduling-Execution-Re-scheduling paradigm.

\section*{Acknowledgment}
Thanks to PPIO Cloud Computing (Shanghai) Co., Ltd. for providing the system platform and original logs. This work was supported in part by China NSFC through grant No. 62072332 and China NSFC (Youth) through grant No. 62002260; in part by the Tianjin Xinchuang Haihe Lab under Grant No.22HHXCJC00002.

% \section*{References}

% Please number citations consecutively within brackets \cite{b1}. The 
% sentence punctuation follows the bracket \cite{b2}. Refer simply to the reference 
% number, as in \cite{b3}---do not use ``Ref. \cite{b3}'' or ``reference \cite{b3}'' except at 
% the beginning of a sentence: ``Reference \cite{b3} was the first $\ldots$''

% Number footnotes separately in superscripts. Place the actual footnote at 
% the bottom of the column in which it was cited. Do not put footnotes in the 
% abstract or reference list. Use letters for table footnotes.

% Unless there are six authors or more give all authors' names; do not use 
% ``et al.''. Papers that have not been published, even if they have been 
% submitted for publication, should be cited as ``unpublished'' \cite{b4}. Papers 
% that have been accepted for publication should be cited as ``in press'' \cite{b5}. 
% Capitalize only the first word in a paper title, except for proper nouns and 
% element symbols.

% For papers published in translation journals, please give the English 
% citation first, followed by the original foreign-language citation \cite{b6}.

\bibliographystyle{IEEEtran}
\bibliography{reference}

% \begin{thebibliography}{00}
% \bibitem{b1} G. Eason, B. Noble, and I. N. Sneddon, ``On certain integrals of Lipschitz-Hankel type involving products of Bessel functions,'' Phil. Trans. Roy. Soc. London, vol. A247, pp. 529--551, April 1955.
% \bibitem{b2} J. Clerk Maxwell, A Treatise on Electricity and Magnetism, 3rd ed., vol. 2. Oxford: Clarendon, 1892, pp.68--73.
% \bibitem{b3} I. S. Jacobs and C. P. Bean, ``Fine particles, thin films and exchange anisotropy,'' in Magnetism, vol. III, G. T. Rado and H. Suhl, Eds. New York: Academic, 1963, pp. 271--350.
% \bibitem{b4} K. Elissa, ``Title of paper if known,'' unpublished.
% \bibitem{b5} R. Nicole, ``Title of paper with only first word capitalized,'' J. Name Stand. Abbrev., in press.
% \bibitem{b6} Y. Yorozu, M. Hirano, K. Oka, and Y. Tagawa, ``Electron spectroscopy studies on magneto-optical media and plastic substrate interface,'' IEEE Transl. J. Magn. Japan, vol. 2, pp. 740--741, August 1987 [Digests 9th Annual Conf. Magnetics Japan, p. 301, 1982].
% \bibitem{b7} M. Young, The Technical Writer's Handbook. Mill Valley, CA: University Science, 1989.
% \end{thebibliography}
% \vspace{12pt}
% \color{red}
% IEEE conference templates contain guidance text for composing and formatting conference papers. Please ensure that all template text is removed from your conference paper prior to submission to the conference. Failure to remove the template text from your paper may result in your paper not being published.

\end{document}